\documentstyle[prl,aps,twoside,epsfig]{revtex}
\input epsf
\evensidemargin=\oddsidemargin
\renewcommand{\thefootnote}{\fnsymbol{footnote}}

\def\del{\partial }
\def\O{{\cal O}}

\def\l{\ell}

\begin{document}
\draft
\twocolumn[\hsize\textwidth\columnwidth\hsize\csname@twocolumnfalse%
\endcsname
\title{Testing Anomalous Gauge Couplings of the Higgs Boson \\
via Weak-Boson Scatterings at the LHC}
\bigskip
\author{Bin Zhang}
\address{Department of Physics, Tsinghua University, Beijing 100084, China.}
\author{Yu-Ping Kuang}
\address{China Center of Advanced Science and Technology (World
Laboratory), P.O.Box 8730, Beijing 100080, China;\\
Department of Physics, Tsinghua University, Beijing
100084, China.\footnote{Mailing address.}}
\author{Hong-Jian He}
\address{Center for Particle Physics,
University of Texas at Austin, Austin, Texas 78712, U.S.A.}
\author{C.--P. Yuan}
\address{Department of Physics and Astronomy,
Michigan State University, East Lansing, MI 48824, U.S.A.}
\bigskip\bigskip
\date{TUHEP-TH-01127,~UTHEP-03-13,~MSUHEP-030103}
\maketitle
\begin{abstract}
We propose a sensitive way to test the anomalous $HVV$ couplings
$(V=W^\pm,~Z^0)$ of the Higgs boson ($H$), which can arise from
either the dimension-3 effective operator in a nonlinearly
realized Higgs sector or the dimension-6 effective operators in a
linearly realized Higgs sector, via studying the $VV$ scattering
processes at the CERN LHC. The gold-plated pure leptonic decay modes of the
final state weak bosons in the processes \,$pp\to VVjj$\, are
studied. For comparison, we also analyze the constraints from the
precision electroweak data, the expected precision of the
measurements of the Higgs boson production rate, decay width and
branching ratios at the Fermilab Tevatron Run-2 and the CERN LHC, and the
requirement of unitarity of the $S$ matrix. We show that, with
an integrated luminosity of 300\,fb$^{-1}$ and sufficient
kinematical cuts for suppressing the backgrounds, studying the
process $pp\to W^+ W^+ jj \to \l^+ \nu \l^+ \nu jj$ can probe the
anomalous $HWW$ couplings at a few tens of percent level for the
nonlinearly realized Higgs sector, and at the level of
$0.01-0.08$\,${\rm TeV}^{-1}$ for the linearly realized effective
Lagrangian.
\end{abstract}

\vspace{0.2cm}\hspace{1.84cm}
PACS number(s): 14.65.Ha, 12.15.Lk, 12.60.Nz
\bigskip\bigskip\bigskip\bigskip\bigskip
]

\setcounter{footnote}{0}
\renewcommand{\thefootnote}{\arabic{footnote}}

\section{INTRODUCTION}

The standard model (SM) of the electroweak interactions has proven
to be very successful in explaining all the available experimental
data at the scale $\lesssim \O(100)$\,GeV. However, the mechanism
of the electroweak symmetry breaking (EWSB) remains one of the
most profound puzzles in particle physics. The Higgs sector of the
SM suffers the well-known problems of triviality \cite{trivi} and
unnaturalness \cite{unnatural}, therefore there has to be new
physics beyond the SM above certain high energy scale $\Lambda$.
Within the SM formalism, the precision electroweak data favors a
light Higgs boson. If a light Higgs boson candidate $H$
is found in future collider experiments, such as the Run-2 of
the Fermilab Tevatron, a 2\,TeV $p {\bar p}$ collider, 
or the CERN Large Hadron Collider (LHC), 
a 14\,TeV $p p$ collider, the
next important task is to experimentally measure the gauge
interactions of this Higgs scalar and explore the nature of the
EWSB mechanism. The detection of the anomalous gauge couplings of
the Higgs boson will point to new physics beyond the SM
underlying the EWSB mechanism.

Although the correct theory of new physics is not yet clear, the
effect of any new physics at energy below the cutoff scale
$\Lambda$ can be parametrized as effective interactions in an
effective theory whose particle content is the same as the SM.
This provides a model-independent description, and the anomalous
couplings relative to that of the SM reflect the effect of 
new physics. In the present study, we
assume that all possible new particles other than the lightest
Higgs boson $H$ are heavy and around or above the scale
$\Lambda$, so that only $H$ is relevant to the effective theory.
Therefore, testing the effective anomalous gauge couplings of the
Higgs boson can discriminate the EWSB sector of the new physics model 
from that of the SM. It has been pointed out that sensitive
tests of the anomalous $HVV$ couplings (with $V=W^\pm,~Z^0$) can
be performed via Higgs boson productions at the future high energy
$e^+e^-$ linear colliders (LC) \cite{HZZ,BHLMZ,tdr}. The tests of the
anomalous $HVV$ couplings at hadron colliders via the decay mode
$H\to\gamma\gamma$ and $H\to \tau\tau$ have also been studied, and
the obtained sensitivities are lower than that at the LC
\cite{HgammagammaLHC,HgammagammaTevatron,PRZ}.

In this paper, following our recent proposal\,\cite{PLB-WW}, we
study an additional way to test the anomalous $HVV$ couplings via
the weak-boson scatterings at the LHC. We shall show that at the
LHC, rather sensitive tests of the anomalous $HVV$ couplings can
be obtained by measuring the cross sections  of the longitudinal
weak-boson scattering, $V_LV_L\to V_LV_L~(V_L=W^\pm_L,~Z^0_L)$,
especially $W^+_LW^+_L\to W^+_LW^+_L$. The scattering amplitude
contains two parts: (i) the amplitude $T(V,\gamma)$ related only
to the electroweak gauge bosons as shown in Fig.\,1(a), and (ii)
the amplitude $T(H)$ related to the Higgs boson as shown in
Fig.\,1(b). At high energies, both $T(V,\gamma)$ and $T(H)$
contain a piece increasing with the center-of-mass energy ($E$) as
$E^2$ in the nonlinear realization and as $E^4$ in the linear
realization. 
In the SM, though the individual diagram in Fig.\,1(a) may contribute 
an $E^4$-dependent piece, the sum of all diagrams in Fig.\,1(a) can have
at most $E^2$-dependent contribution, which can be easily verified 
by an explicit calculation. 
Furthermore, the $HVV$ coupling constant
in the SM is fixed to be the same as the non-Abelian gauge
self-coupling of the weak bosons. This causes the two
$E^2$-dependent pieces to precisely cancel with each other in the sum of 
$T(V,\gamma)$ and $T(H)$, resulting in the expected $E^0$-behavior for
the total amplitude, as required by the unitarity of the SM
$S$ matrix. If the $HVV$ couplings are anomalous due to the effect of new
physics above the cutoff scale $\Lambda$, the total amplitude
of $VV$ scatterings can grow as $E^2$ or $E^4$ in the high energy regime. 
Such deviations from the $E^0$ behavior of the SM
amplitude can provide a rather sensitive test of the anomalous $HVV$
couplings in high energy $VV$ scattering experiments.  This type
of tests require neither the detection of the Higgs boson
resonance nor the measurement of the Higgs boson decay branching
ratios, and is thus of special interest.  If the anomalous $HVV$
coupling associated with the new physics effect is rather large, the
total decay width of $H$ may become so large that $H$ cannot be detected
as a sharp resonance\,\cite{broadH} and therefore escapes the
detection when scanning the invariant mass of its decay products
around $m_H$. In that case, can we tell whether or not there exists a
sub-TeV Higgs boson? The answer is yes. It can be tested by
carefully studying the scatterings of weak gauge bosons in the TeV
region\,\cite{kuang_he_cp}. Furthermore, if the new physics causes
a rather small $HVV$ coupling (much below the value of the SM
$HVV$ coupling), the production rate of a light Higgs boson can
become so small that it escapes the detection when the
experimental measurement is taken around the Higgs boson mass
scale. However, as mentioned above, the scattering of $V_LV_L$ has
to become large in the TeV region accordingly.  Therefore, this
makes it important to study the $V_LV_L$ scattering in the TeV
regime even in the case where a light Higgs boson exists in the
EWSB sector. In Sec.\,IV, we show that this type of test of the
anomalous $HVV$ couplings can be more sensitive than those obtained
from studying the on-shell Higgs boson production and decays at the LHC
\cite{HgammagammaLHC,HgammagammaTevatron,PRZ}, as well as the  
constraints derived from the precision electroweak data
and the requirement of the unitarity of the $S$ matrix. Therefore,
 studying the $V_LV_L$ scatterings at the LHC is not only
important for probing the strongly interacting electroweak
symmetry breaking sector without a light Higgs boson,
but also valuable for sensitively testing of the anomalous gauge
interactions of the Higgs boson when a light Higgs scalar exists
in the mass range $115-300$\,GeV. This further supports the
``no-lose'' theorem\,\cite{mike} for the LHC to decisively probe
the EWSB mechanism.

This paper is organized as follows. In Sec.\,II, we briefly sketch
a few key points related to the
calculations of the $VV$ scatterings discussed in this paper.
In Sec.\,III, we systematically study the test of the
anomalous $HVV$ coupling from the dimension-3 operator in a
nonlinearly realized Higgs sector\,\cite{CK}, as an extension of
\cite{PLB-WW}.
The test of the anomalous $HVV$ couplings from the dimension-6 operators
in the linearly realized Higgs sector\,\cite{linear,G-G}
is studied in Sec.\,IV.  In both cases, the
constraints on the anomalous $HVV$ couplings from the precision
electroweak data and the requirement of the unitarity of the
$S$ matrix are also discussed. Section\,V summarizes our concluding remarks.


\section{ $VV$ SCATTERINGS}

As mentioned in the previous section, we will take the enhanced
$VV$ scatterings as the signals for testing the anomalous $HVV$
couplings. We choose the gold-plated pure leptonic decay modes of
the final state weak bosons as the tagging modes, this will avoid
the large hadronic backgrounds at the LHC. Even in this case,
there are still several kinds of backgrounds to be eliminated,
namely the {\it electroweak (EW) background}, the {\it QCD
background}, and the {\it top quark background} studied in
Refs.\,\cite{WW94,WW95}. At the LHC, the initial state $V$'s in the
$VV$ scattering are emitted by the quarks in the protons. As
Refs.\cite{WW94,WW95,WW90,Yuan} pointed out, the QCD background
can be greatly suppressed by tagging a forward-going jet (the
out-going quark after emitting the $V$). This forward-jet tagging
will also suppress the transverse component ($V_T$) of the initial
$V$, so that the initial state $V$ will be
essentially $V_L$. Following Refs.\cite{WW94,WW95}, we impose, in
addition to {\it forward jet tagging}, the requirement of {\it
vetoing the central jet} to further suppress the QCD background
and the top quark background.
The EW background can be suppressed by {\it detecting isolated leptons
with large transverse momentum
 in the central rapidity region}, especially requiring the two
final state leptons to be nearly back to back. These leptonic cuts also
suppress the $V_T$ contributions in the final state.
Moreover, choosing the decay leptons in
the central rapidity region also avoids the
collinear divergence in the diagrams exchanging a photon in $T(V,\gamma)$.

In this paper, we shall calculate the complete tree level
contributions to the processes
\begin{eqnarray}                        
pp\to VVjj, \label{VVjj}
\end{eqnarray}
where $j$ is the forward jet. We shall impose all the cuts
mentioned above, and use the updated CTEQ6L \cite{cteq6} parton
distribution functions for the distributions of quarks in the
protons. We also take into account the effect of the
width of the weak boson in calculating the helicity amplitudes.

References \cite{WW94,WW95} studied various strongly interacting models
which do not consist of a light Higgs boson and the $V_LV_L \to
V_LV_L$ scattering amplitudes are largely enhanced (by powers of
$E^2$) in the high energy region. The signal amplitudes were
calculated in the effective $W$ approximation (EWA) \cite{EWA}
because the $V_LV_L \to V_LV_L$ scattering amplitudes predicted by
those models violate the unitarity condition of a $2 \to 2$
scattering matrix so that a full $2 \to 4$ process $pp \to V_L V_L
jj$ could not be reliably calculated to predict the signal event
rates in the TeV region. Instead, the $V_LV_L \to V_LV_L$
scattering amplitudes were properly unitarized before they were
convoluted with the $W$-boson luminosities obtained from the EWA
to predict the signal event rates. With this method of
calculation, it was shown that after imposing the kinematic
requirements discussed above, the backgrounds are reasonably
suppressed relative to the signals, so that the $V_LV_L$
scatterings signals can be effectively extracted.

As to be discussed in Secs.\,III and VI, in our present case, the
Higgs boson is light, and the new physics effect is assumed to only
modify the effective operators in either a nonlinearly 
or a linearly realized Higgs sector. In the nonlinearly
realized case, the $V_LV_L$ scattering amplitudes with a not too
small anomalous $HVV$ couplings are also enhanced in the high
energy region due to the $E^2$ behavior of the amplitudes. Hence,
we may apply the same methodology as proposed in Ref. \cite{WW95}
for calculating the signal rate. However, in this work, we shall
not choose using the EWA. Instead, we shall compute the full $pp
\to VV jj$ cross sections at the tree level, cf.
Eq.\,(\ref{VVjj}), which is justified as long as the LHC
sensitivity to the size of the anomalous $HVV$ coupling is smaller 
than that required to render the unitarity of the 
$V_LVL \to V_LV_L$ partial wave amplitudes.
As to be shown in Secs.\,IV and VI,  this is
indeed the case and the backgrounds can be reasonably suppressed
relative to the signals by imposing the same kinematical cuts
suggested in Ref.\,\cite{WW95}. On the contrary, in the case of
the SM (without anomalous $HVV$ couplings), all the $V_LV_L$,
$V_TV_L$, and $V_TV_T$ amplitudes behave as $E^0$ rather than
$E^2$. Because the probability of finding a transverse vector
boson with high momentum is much larger than a
longitudinal vector boson  \cite{EWA}, the contribution of
$V_LV_L \to V_LV_L$ to the production rate of $pp \to VV jj$ in
the TeV region is small as compared to the $V_TV_L$ and $V_TV_T$
contributions. Therefore, although the imposed kinematics cuts can
effectively suppress the QCD and the top quark backgrounds, they
leave a considerable EW background (mainly originated from the
$V_TV_L$ and $V_TV_T$ contributions) that dominates the $VV$
scattering in the SM. Throughout this paper, we will call these
$VV$ scattering contributions the remaining SM electroweak
backgrounds after imposing the kinematical cuts. We shall do
a complete tree-level calculation for both the signal and the
background processes with the improvement on
the simulation of the kinematical distributions of the decay leptons, 
and 
the cross section calculation with the updated parton distribution 
functions~\cite{cteq6} and the
inclusion of the vector boson width. 
Furthermore, we do not apply the EWA \cite{EWA} in our calculations, 
hence, our results are not identical to 
those given in Ref.\,\cite{WW95}, even for the SM case.

In the case of a linearly realized effective Lagrangian, the
physics consideration is quite different. As to be shown in
Sec.\,VI, the $V_LV_L \to V_LV_L$ scattering amplitudes can have
not only the $E^2$ but also the $E^4$ contributions depending on
the process and operator under consideration.  Furthermore, both
the $V_TV_T \to V_LV_L$ and $V_LV_T \to V_LV_T$ scattering
amplitudes can also have the $E^2$ contributions, which should be
undoubtedly counted as part of the {\it signal} rates because
those contributions are absent in the SM. Because the luminosity
of $V_T$ is larger than the $V_L$, the contributions from $V_TV_T$
and $V_TV_L$ scatterings cannot be ignored unless the $E^4$
contributions dominate the $E^2$ contributions which occur only
when both the energy $E$ and the anomalous couplings become large.
It implies that including only the $V_LV_L \to V_LV_L$ contribution 
with the EWA, as done in Ref.\,\cite{WW95}, is not
adequate in this case, and a full $2 \to 4$ calculation should be
used to calculate the signal rates. Although it is possible to
apply the EWA to include also the $V_TV_T$ and $V_LV_T$
contributions\,\cite{he-zerwas} with jet tagging efficiencies (for
$V_L$ and $V_T$, separately) extracted from the study done in
Ref.\,\cite{WW95}, we choose to perform a full $pp \to VV jj$
tree-level calculation which is justified as long as the
unitarity condition is satisfied.
Nevertheless, as to be discussed in the end of Sec.~VI, 
we shall apply the EWA, folded with the $VV \to VV$ 
scattering amplitudes, to check the high energy behavior
of the full $2 \to 4$ amplitudes which are also
verified to be gauge invariant.

\section{TESTING ANOMALOUS HVV COUPLING\\
         FROM DIMENSION-3 OPERATOR}

\subsection{The Anomalous HVV Coupling from Dimension-3 Operator}

It is known that there is no anomalous $HVV$ coupling arising from
the dimension-3 and dimension-4 gauge invariant operators in
the linearly realized effective Lagrangian\,\cite{linear,G-G}.
Here, we consider the nonlinearly realized
Higgs sector formulated in Ref.\,\cite{CK}.
In this nonlinear formalism,
the effective Lagrangian below the cutoff scale $\Lambda$
contains the Higgs field $H$ transforming as a weak singlet, the would-be
Goldstone boson field ${\overrightarrow \omega}$, and the
electroweak gauge boson fields, and it respects the electroweak gauge symmetry,
charge conjugation {\tt C} and parity {\tt P}, and the
custodial $SU(2)_c$ symmetry.
Up to dimension-4, the effective Lagrangian is given by\,\cite{CK}
\begin{eqnarray}                            
{\cal L}&=&-\frac{1}{4}{\overrightarrow
W}_{\mu\nu}\cdot{\overrightarrow
W}^{\mu\nu}-\frac{1}{4}B_{\mu\nu}B^{\mu\nu}\nonumber\\
&&+\frac{1}{4}(v^2+2\kappa vH+\kappa^\prime H^2){\rm
Tr}(D_\mu\Sigma^\dagger D^\mu\Sigma)\nonumber\\
&&+\frac{1}{2}\del_\mu H\del^\mu H
-\frac{m_H^2}{2}H^2-\frac{\lambda_3 v}{3!}H^3+\frac{\lambda_4}{4!}H^4,
\label{Lagrangian}
\end{eqnarray}
where $\overrightarrow W_{\mu\nu}$ and $B_{\mu\nu}$ are field
strengths of the electroweak gauge fields, $v\simeq 246$\,GeV is the
vacuum expectation value breaking the electroweak gauge symmetry,
$(\kappa,\,\lambda_3)$ and
$(~\kappa^\prime,\,\lambda_4)$ are
dimensionless coupling constants from the dimension-3 and
dimension-4 operators, respectively.
In Eq.\,(\ref{Lagrangian}), we have defined
\begin{eqnarray}                               
&&\Sigma=\exp{\bigg(\frac{i{\overrightarrow\tau}\cdot{\overrightarrow\omega}}
{v}\bigg)},\nonumber\\
&&D_\mu\Sigma=
\partial_\mu\Sigma+ig\frac{\overrightarrow\tau}{2}\cdot{\overrightarrow
W}_\mu\Sigma -ig'B_\mu\Sigma\frac{\tau_3}{2} \, ,
\end{eqnarray}
in which the Pauli matrix $\tau_i$ is normalized as
${\rm Tr}(\tau_i \tau_j)=2 \delta_{ij}$,
and $g$ and $g'$ are the $SU(2)$ and $U(1)$ gauge coupling
constants, respectively.
The SM corresponds to
$\kappa=\kappa^\prime=1$ and
$\displaystyle\lambda_3=\lambda_4 =\lambda={3 m_H^2}/{v^2}$.

We note that at the tree level, only the dimension-3 operator
$\frac{1}{2}\kappa vHD_\mu\Sigma^\dagger D^\mu\Sigma$ in
Eq.\,(\ref{Lagrangian}) contains the anomalous $HVV$ coupling that
can contribute to the $VV$ scatterings (cf. Fig.\,1). Therefore,
$VV$ scatterings can test this dimensionless coupling $\kappa$,
and $\Delta\kappa\equiv\kappa -1$ measures the deviation from the
SM value $\kappa =1$.

\subsection{Precision Constraints on the Coupling ${\mathbf\kappa}$}

Equation (\ref{Lagrangian}) shows an important difference between the
nonlinearly and the linearly realized Higgs sectors. The nonlinear
formalism allows new physics to appear in the effective operators
with dimension $\leq 4$ whose coefficients are not necessarily
suppressed by the cutoff scale $\Lambda$. Hence the gauge
couplings of Higgs boson to weak bosons can naturally deviate from
the SM values by an amount at the order of $\lesssim {\cal O}(1)$,
as suggested by the naive dimensional analysis\,\cite{NDA}.

In the unitary gauge, the anomalous couplings of Higgs boson 
to gauge bosons relevant to
the precision oblique parameters $(S,\,T,\,U)$ \cite{PT} are
\begin{eqnarray}                               
\displaystyle
{[4(\kappa-1)vH +2(\kappa'-1)H^2] \over v^2}
\left[{m_W^2}W^+_\mu W^{-\mu}
 +\frac{1}{2} {m_Z^2} Z_\mu Z^\mu \right] .
\nonumber
\end{eqnarray}
When calculating radiative corrections using the effective
Lagrangian (\ref{Lagrangian}), it is generally necessary to
introduce higher dimensional counter terms to absorb the new
divergences arising from the loop integration. There are in
principle three next-to-leading order (NLO) counter terms to
render the $(S,~T,~U)$ parameters finite at the one-loop level,
namely \cite{App}
\begin{eqnarray}                         
\label{eq:L018}
\displaystyle {\cal L}^{(2)^\prime} &=&~
\ell_0 \frac{v^2}{16\pi^2}
\frac{1}{4}
[{\rm Tr}{\cal T} (D_\mu{\Sigma}){{\Sigma}}^\dagger]^2,
\nonumber
\\[2.5mm]
{\cal L}^{(4)}_1 &=&~ \displaystyle
\ell_1 \frac{v^2}{\Lambda^2} gg'
{\rm Tr}[{\bf B}_{\mu\nu}\Sigma^\dagger {\bf W}^{\mu\nu}\Sigma]\,,
\\[2.5mm]
{\cal L}^{(4)}_8 &=&~
\displaystyle
\ell_8 \frac{v^2}{\Lambda^2}\frac{g^2}{4}
[{\rm Tr}({\cal T} {\bf W}_{\mu\nu})]^2 \, ,
\nonumber
\end{eqnarray}
whose coefficients $(\ell_0,\ell_1,\ell_8)$ correspond to the
oblique parameters $(T,\,S,\,U)$, respectively. Here ${\bf
W}_{\mu\nu} ={\overrightarrow W}_{\mu\nu}\!\cdot\!
{\overrightarrow  \tau}/2$, ${\bf B}_{\mu\nu}
={B}_{\mu\nu}\tau_3/2$, and ${\cal
T}\equiv\Sigma\tau_3\Sigma^\dagger$. Comparing to ${\cal
L}^{(4)}_1$, ${\cal L}^{(4)}_8$ is of the same
dimension, but with two new
$SU(2)_c$-violating operators $\cal T$, so that we expect
$\ell_8/\ell_1 \sim 1/16\pi^2 \sim 10^{-2} \ll 1$. This generally
leads to \,$U\ll S,\,T$\,.\, To estimate the contribution of loop
corrections, we invoke a naturalness assumption that no fine-tuned
accidental cancellation occurs between the leading logarithmic
term and the constant piece of the counter terms. Thus, the
leading logarithmic term represents a reasonable estimate of the
loop corrections. This approach is commonly used in the literature
for estimating new physics effects in effective
theories\,\cite{eg}. It is straightforward to compute the
radiative corrections to $S,\,T,$ and $U$, arising from the $HVV$
anomalous couplings, using dimensional regularization in the
$\overline{\rm MS}$ scheme and keeping only the leading logarithmic
terms. After subtracting the SM Higgs contributions ($\kappa=1$)
with  the reference value $m_H^{\rm ref}$, we find
\begin{eqnarray}                        
\label{eq:STUH}
\hspace*{-5mm}
&&\Delta S = \displaystyle
\frac{1}{6\pi}[\ln\frac{m^{~}_H}{\,m_H^{\rm ref}}
     -(\kappa^2\!-\!1)\ln\frac{\Lambda}{\,m^{~}_H} ]\, ,\!\nonumber\\
&&\Delta T = \displaystyle
\frac{3}{8\pi c_{\rm w}^2}
[-   \ln\frac{m^{~}_H}{\,m_H^{\rm ref}}
+(\kappa^2\!-\!1)\ln\frac{\Lambda}{\,m^{~}_H}],\nonumber\\
&&\Delta U = \displaystyle 0,
\end{eqnarray}
where the terms containing $\ln\Lambda$ represent the genuine new
physics effect arising from physics above the cutoff scale
$\Lambda$\,\cite{BL}. Note that at the one-loop order, the
coupling $\kappa'$ has no contribution to the $S$, $T$ and $U$
parameters. In the SM ($\kappa=1$), an increase of the Higgs mass
will increase $\Delta S$ and decrease $\Delta T$. Choosing the
reference Higgs mass $m_H^{\rm ref}$ to be $m^{~}_H$, we may
further simplify Eq.\,(\ref{eq:STUH}) as
\begin{eqnarray}                       
&&\Delta S = \displaystyle
  -\frac{\,\kappa^2-1\,}{6\pi}\ln\frac{\Lambda}{\,m^{~}_H},\nonumber\\
&&\Delta T = \displaystyle
  +\frac{\,3(\kappa^2-1)\,}{8\pi c_{\rm w}^2}\ln\frac{\Lambda}{\,m^{~}_H},
\nonumber\\
&&\Delta U = \displaystyle 0.
\label{eq:STUf}
\end{eqnarray}
For a given value of $\Lambda$ and $m^{~}_H$, we find 
  $\Delta S < 0$ and $\Delta T> 0$ for $|\kappa| >1$.
This pattern of radiative corrections allows a relatively heavy
Higgs boson to be consistent with the current precision data [cf.
Fig. 2(a)]. Furthermore, from Eq.\,(\ref{eq:STUf}) we see that the
loop contribution from $\kappa\neq 1$ results in a sizable ratio
of \,$\Delta T/\Delta S = -9/(4c_w^2) \approx -3$\,.

Within the framework of the SM, the global fit to
the current precision electroweak data favors a
light Higgs boson with a central value $m^{~}_H=83$\,GeV
(significantly below the CERN $e^+e^-$ collider LEP2 direct search limit $m^{~}_H >
114.3$\,GeV \cite{PDG}) and a 95\%\,C.L. limit, $\,32\,{\rm
GeV}\,\leq m^{~}_H \leq 192$ GeV, on the range of the Higgs boson
mass. However, it was recently shown that in the presence of new physics, 
such a bound can be substantially
relaxed\,\cite{seesaw,bagger,HPS,snow,Mike}. The latest NuTeV data
was not included in the above analysis. If we include the NuTeV
data, the value of the minimum $\chi^2$ of the global fit
increases substantially (by $8.7$), showing a poor quality of the
SM fit to the precision data (this fit also gives a similar
central value \,$m^{~}_H=85$\,GeV\, and 95\%\,C.L. mass range
$\,33\,{\rm GeV}\,\leq m^{~}_H \leq 200$\,GeV).  We also find a
similar increase of $\chi^2$ (by 8.9) in the $\Delta S-\Delta T$
fits, which implies that the NuTeV anomaly may not be explained by
the new physics effect from the oblique parameters $(\Delta S,\,\Delta
T)$ alone. As the potential problems with the NuTeV analysis are
still under debate\,\cite{NuTeVx}, we will not include the NuTeV
data in the following analysis. [But for comparison,  we have also
displayed a $95\%$ C.L. contour (dotted curve) from the fit
including the NuTeV anomaly in Fig. 2(a).] In Fig. 2(a), we show
the $\Delta S-\Delta T$ bounds (setting $m_H^{\rm ref} =
100$\,GeV) derived from the global fit with the newest updated
electroweak precision data\,\cite{elgapp,data-new}. Furthermore,
for $m_H^{\rm ref} = 115\,(300)$\,GeV and $\Delta U=0$, we find
that the global fit gives
\begin{eqnarray}            
&&\Delta S=0.01\,(-0.07)\pm 0.09,\nonumber\\
&&\Delta T=0.07\,(0.16)\pm 0.11.
\label{ST}
\end{eqnarray}
In the same figure, we also plot the SM Higgs boson contributions
to $\Delta S$ and $\Delta T$ with different Higgs masses. Figure
2(b) shows that the upcoming measurements of the $W^\pm$ mass
($m_W$) and top mass ($m_t$) at the Tevatron Run-2 can
significantly improve the constraints on new physics via the
oblique corrections, in which we have input the current Run-1
central values of $(m^{~}_W,\,m^{~}_t)$ and the expected errors of
$(m^{~}_W$ and $m^{~}_t)$ from the planned Run-2 sensitivity of
$20$\,MeV and $2$\,GeV, respectively.

Using the allowed range of $(\Delta S,\,\Delta T)$ in Fig.\,2(a),
we can further constrain the new physics scale $\Lambda$ as a
function of the anomalous coupling $\Delta\kappa$ for some typical
values of $m_H^{\rm ref}=m^{~}_H$ [cf. Eq. (\ref{eq:STUf})].
Figure 3 depicts these constraints. With Fig.\,3, we can
alternatively  constrain the range of $\Delta\kappa$ for a given
value of $(\Lambda,\,m^{~}_H)$, which is summarized in Table\,I.
Figure 3 and Table\,I show that for $\,m^{~}_H\agt 250-300$\,GeV,
the $\,\Delta\kappa < 0\,$ region is fully excluded, while a
sizable $\,\Delta\kappa > 0\,$ is allowed so long as $\Lambda$ is
relatively low. Moreover, for $\,m^{~}_H\agt 800$\,GeV, the region
$\,\Lambda<1.1$\,TeV is excluded. For the range of $\,m^{~}_H\agt
250-300$\,GeV, the preferred values of $\Delta\kappa > 0$ require
the $HW^+W^-$ and $HZZ$ couplings to be stronger than those in the
SM. In this case, the direct production rate of the Higgs boson,
via either the Higgs-strahlung or the $VV$ fusions in high energy
collisions, should raise above the SM rate. On the other hand, for
$m^{~}_H\lesssim 250$\,GeV, the direct production rate of the
Higgs boson can be smaller or larger than the SM rate depending on
the sign of $\Delta \kappa$. Hence, when $\Delta \kappa < 0$, a
light nonstandard Higgs boson may be partially {\it hidden} by
its large SM background events. However, in this situation, the
new physics scale $\Lambda$ will be generally low. For instance,
when $m^{~}_H=115$\,GeV, a negative $\Delta\kappa =
-0.15\,(-0.28)$ already forces $\Lambda\leq 1\,(0.4)$\,TeV.
Finally, we comment that, for a certain class of models, new
physics effects may also be induced by extra heavy fermions such
as in the typical top-seesaw models with new vector-like
fermions\,\cite{seesaw,DSB} or models with new chiral
families\,\cite{HPS}. In these models, there can be generic
positive contributions to $\Delta T$, so that the $\Delta\kappa <
0$ region may still be allowed for a relatively heavy Higgs boson,
but such possibilities are very model-dependent. In our current
effective theory analysis, we consider the bosonic $HVV$ couplings
as the dominant contributions to the oblique parameters, and
assume that other possible anomalous couplings (such as the
deviation in the gauge interactions of $tbW/t\bar{t}Z$
\cite{ehab}) can be ignored. However, independent of these
assumptions, the {\it most decisive test} of the $HVV$ couplings
can come from the direct measurements via Born-level processes at
the high energy colliders, which is the subject of the next two
sections.

\subsection{Constraints on $\bf\kappa$ from Unitarity Requirement}

Before concluding this section, we discuss the possibility of
unitarity violation in the scattering process \,$pp\to W^+W^+
jj\to \ell\nu \ell \nu jj$ for $\kappa \ne 1$, whose leading
contribution comes from the subprocess $W^+_L W^+_L \to W^+_L
W^+_L$ when the initial $W_L^+$ bosons are almost collinearly
radiated from the incoming quarks or antiquarks. The scattering
amplitude of $W^+_L W^+_L \to W^+_L W^+_L$ contributes to the
isospin $I=2$ channel, and in the high energy region ($E^2\gg
m_W^2, m^2_H$), it is dominated by the leading
$E^2$-contributions, and
\begin{equation}
T[I=2] ~\simeq~ (\kappa^2-1)\displaystyle\frac{E^2}{v^2} \,.
\end{equation}
Using the partial-wave analysis, the $s$ wave amplitude
~$a^{~}_{I,J=2,0}$ is found to be 
\begin{eqnarray}                        
a^{~}_{20} ~\simeq~\displaystyle (\kappa^2-1)\frac{E^2}{\,16\pi v^2\,},
\end{eqnarray}
where $E=M_{VV}$ is the invariant mass of the vector boson pair.
The unitarity condition for this channel is\footnote{Another
convention in the literature \cite{WW94,WW95} reads, $\,|{\rm
Re}~a^{~}_{20}| < 1/{2}$\,,\, for $a^{~}_{20}
~\simeq~\displaystyle (\kappa^2-1){E^2}/(32\pi v^2)$. In this
convention, the partial wave amplitude of the elastic scattering
$W^+W^+ \to W^+W^+$, beyond the tree level, satisfies the relation
$\,{\rm Im}~a = |a|^2$.}
$\,|{\rm Re}~  
a^{~}_{20}| < {2!}/{2}=1$\,,\, in which the factor $2!$ is due to
the identical particles ($W^+W^+$) in the final state. This results in a
requirement that 
\begin{eqnarray}                      
\label{eq:UB-K}
\displaystyle
\sqrt{1-\frac{\,16\pi v^2\,}{E^2}}~<~|\kappa|~<~\sqrt{1+\frac{\,16\pi v^2\,}
{E^2}}.
\end{eqnarray}
For example,
\begin{eqnarray}                       
&~~~~~~~~|\kappa| < 3.6 \,,&~~~~~~{\rm for}~~~  E=0.5~{\rm TeV},\nonumber\\
&0.5 < |\kappa| < 1.3  \,,&~~~~~~{\rm for}~~~
E=2~{\rm TeV},\nonumber\\
& 0.8 < |\kappa| < 1.2  \,,&~~~~~~{\rm for}~~~
 E=3~{\rm TeV}. \label{Ubound}
\end{eqnarray}
As to be shown in the next section, the expected sensitivity of
the LHC in determining $\Delta\kappa$ [cf. Eq.~(\ref{precision})]
is consistent with the unitarity limit since the typical
invariant mass of the $W^+W^+$ pair, after the kinematic cuts,
falls into the range \,$500\,{\rm GeV} \leq E \leq 2\sim 3$\,TeV.
The contributions from higher invariant mass values are severely
suppressed by parton luminosities\,\cite{WW95,power}, and are
thus negligible.

\section{Testing ${\mathbf \kappa}$ via VV Scatterings at the LHC}

Knowing the above constraints, we turn to the analysis of testing
$\kappa$ via $VV$ scatterings at the LHC.
Following the procedures described in Sec.\,II,
we calculate the tree-level cross sections of the scattering processes
\begin{eqnarray}                 
&&pp\to Z^{}_LZ^{}_L jj\to \l^+\l^-\l^+\l^-jj,\l^+\l^-\nu\bar{\nu}jj,
\nonumber\\
&&pp\to W^+_LW^-_L jj\to \l^+\nu \l^-\bar{\nu}jj,\nonumber\\
&&pp\to W^+_LW^+_L jj\to \l^+\nu \l^+\nu jj,\nonumber\\
&&pp\to W^-_LW^-_L jj\to \l^-\bar{\nu}\l^-\bar{\nu} jj,\nonumber\\
&&pp\to Z^{}_LW^+_L jj\to \l^+\l^-\l^+\nu jj,\nonumber\\
&&pp\to Z^{}_LW^-_L jj\to \l^+\l^-\l^-\bar{\nu} jj.
\label{channels}
\end{eqnarray}
\null\noindent

In our numerical calculations, we shall take the same kinematic
cuts as those proposed in Ref.\cite{WW95} to suppress the backgrounds 
discussed in Sec. II. It has been shown in Table II of
Ref.\cite{WW95} that, after imposing the kinematical cuts, the sum of the
cross sections of the QCD background and the top quark background
is smaller than the cross section of the remaining electroweak
background by one order of magnitude. 
Thus, the EW background needs to be studied in more detail.
For that, we have examined the 
polarization of the final state $W^+$ bosons.
In Table II, we list the cross section of 
$pp \to W^+_{\lambda_1} W^+_{\lambda 2} jj$ at the LHC,
for various values of $\Delta\kappa$ with $m_H=115$ GeV,
where $W^+_\lambda$ denotes a polarized $W$ boson with 
the polarization index $\lambda=T$ or $L$.
As expected, when $|\Delta\kappa|$ is large,
the $W^+_LW^+_L$ production rate dominates, and
all the SM backgrounds, after the kinematical cuts, are
reasonably suppressed relative to the signal due to 
the $E^2$-dependence of the $W_LW_L$ amplitude.
 However, for $\Delta\kappa=0$ (the SM case), although
the QCD and top quark backgrounds are negligibly small, the EW
background contributed from the $TT$ and $LT$ polarizations are
still quite large as compared to the signal, cf. Table II. 
Therefore, in this
work, we shall take the cross section for $\Delta\kappa=0$ (the
SM case) as the remaining EW background cross section.

In Tables\,III$\--$VI, we list the obtained numbers of events
(including both the signals
and backgrounds) for an integrated luminosity of 300 fb$^{-1}$
with the parameters $115~{\rm GeV}\le m_H\le 300$ GeV and
$-1.0\le\Delta\kappa\le 0.7$. In
general, the number of events for $\kappa>0$ and $\kappa<0$ are
not the same. However, the result of our calculations shows that the
difference between them is very small. Thus we only list the number of
events with $\Delta\kappa\ge -1~(\kappa\ge 0)$ in the Tables. We
see that the most sensitive channel to determine the anomalous
coupling $\Delta\kappa$ is $pp\to W^+W^+jj\to l^+\nu l^+\nu
jj$ (cf. Table III) due to its small background rates, and thus
the weakest kinematic cuts \cite{WW95}.

It is easy to check that the numbers in Tables V and VI, for the
$ZW^\pm$ and $ZZ$ channels, are consistent with the unitarity
bound. Only the $W^+W^-$ channel, cf. Table\,IV, needs further
unitarization \cite{WW94,WW95}. It is expected that after unitarizing
the scattering amplitudes, the corresponding numbers in Table\,IV
will become smaller, and thus this channel is less interesting.
Therefore in this paper we only take the best channel, the
$W^+W^+$ channel (cf. Table\,III), to constrain $\Delta \kappa$.

As discussed above, we shall take the SM events (for
$\Delta \kappa=0$) listed in Table\,III as the intrinsic SM
electroweak background rate in our analysis,
i.e., $N_B\equiv N(\Delta\kappa=0)$.
We can then define the number of signal events 
(for $\Delta\kappa\ne 0$) as
$N_S\equiv N(\Delta\kappa \ne 0)-N_B$. The total statistical
fluctuation is $\sqrt{N_S+N_B}$. To study the potential of the LHC
in distinguishing the $\Delta\kappa \ne 0$ case from the SM, we
also show the deviation of the signal from the total statistical
fluctuation, $N_S/\sqrt{N_S+N_B}$, in the parentheses in
Tables\,III. The values of $N_S/\sqrt{N_S+N_B}$ in Table III show
that the LHC can limit $\Delta\kappa$ to the range
\begin{eqnarray}                    
-0.3<\Delta\kappa<0.2
\label{precision}
\end{eqnarray}
at roughly the $(1-3)\sigma$ level if no anomalous coupling
($\Delta\kappa\ne 0$) effect is detected.

As mentioned in Sec.\,I, if the new physics above the cutoff scale
$\Lambda$ happens to make $\Delta\kappa$ negative and close to
$\Delta\kappa=-1$ ($\kappa \agt 0$), the Higgs production rate may
become so small that the Higgs resonance may be difficult to detect.
However, we see from Tables\,III that the event numbers of $pp\to
W^+W^+ jj\to l^+\nu l^+\nu jj$ for this $\kappa$ value are much
larger than those for the SM, so that we can clearly detect the
$\kappa \agt 0$ effect via these processes without the need of
detecting the resonance of a light Higgs boson. This is the clear
advantage of this type of measurements when the resonant state of
the Higgs boson is difficult to be directly detected experimentally.

\section{Other Possible Tests of ${\mathbf \kappa}$
         from Future Collider Experiments}

There can be other tests of $\kappa$ from future collider
experiments. After discussing a few relevant 
measurements available at the Fermilab Tevatron and the CERN LHC, 
we shall also comment on the potential
of the future LC. They are given below in order. 

\subsection{Associate $HV$ Production}

First, let us consider the associate production of a Higgs boson and
vector boson ($H+V$) at high energy colliders.
The studies at LEP-2 concluded that the mass of the SM Higgs boson has to be
larger than about 114.3 GeV \cite{PDG}. For a non-SM Higgs boson, this lower
mass bound will be different. For example, in the supersymmetric SM, the
$H$-$Z$-$Z$ coupling is smaller than that of the SM by a factor of 
$\sin(\alpha-\beta)$ or $\cos(\alpha-\beta)$, depending on whether
$H$ denotes a light or a heavy CP-even Higgs boson, and the
above-mentioned lower
mass bound is reduced to about 90 GeV \cite{EWWG}. If the mass of
the SM Higgs boson is around 110 GeV, it can be discovered at the
Tevatron Run-2. Assuming an integrated luminosity of 10 fb$^{-1}$,
the number of the expected signal events is about 27 and the
background events about 258, according to the Table~3
 (the most optimal scenario) of Ref.~\cite{agrawal}. Therefore, the $1\sigma$
statistic fluctuation of the total event is $\sqrt{27+258}
\sim 17$. Consequently, at the $1\sigma$ level, $0.6 < |\kappa|
< 1.2$, and at the $2\sigma$ level, $|\kappa| <
1.5$.\footnote{This bound can be improved by carefully examining
the invariant mass distribution of the $b {\bar b}$ pairs.}
Similarly, we estimate the bounds on $|\kappa|$
for various $m_H$ as follows:
\begin{eqnarray}                      
&&m_H({\rm GeV})\hspace{1.7cm} 1\sigma \hspace{2.0cm} 2\sigma\nonumber\\
&&110:
\hspace{1.5cm}0.6 \le |\kappa|\le 1.2,\hspace{0.3cm}  0 \le |\kappa|\le 1.5,
\nonumber\\
&&120:
\hspace{1.5cm}0.4 \le |\kappa|\le 1.4,\hspace{0.3cm} 0 \le |\kappa| \le 1.6,
\nonumber\\
&&130:~
\hspace{1.5cm} {  }0 \le |\kappa|\le 1.5,\hspace{0.6cm} 0 \le |\kappa|\le 1.8,
\label{tevbound}
\end{eqnarray}
We see that the above limits on $\kappa$ are
weaker than that in Eq. (\ref{precision}). Of course, the above
bounds can be further improved at the Tevatron Run-2 by having a
larger integrated luminosity until the systematical error
dominates the statistical error. The same process can also be
studied at the LHC to test the anomalous coupling $\kappa$.
However, because of the much larger background rates at the LHC,
the improvement on the measurement of $\kappa$ via the associate
production of $V$ and Higgs boson is not expected to be
significant.

\subsection{Measuring Total Decay Width of a Higgs Boson}

The anomalous $HVV$ coupling can also be tested from 
measuring the total decay width of the Higgs boson.
 When the Higgs boson mass is larger than twice the
$Z$ boson mass, it can decay into a $Z$ boson pair which
subsequently decay into four muons. This decay channel is labeled as
one of the ``gold-plated'' channels
(the pure leptonic decay modes)
because it is possible to construct the
invariant mass of the $ZZ$ pair in this decay mode with a high
precision, which provides a precision measurement on the total
decay width of the Higgs boson. A detailed Monte Carlo analysis
has been carried out in Ref.~\cite{width} to find out the
uncertainty on the measurement of the Higgs boson width. Assuming
that there is no non-SM decay channel opened and only the
anomalous $HVV$ coupling is important, one can convert the
conclusion from~\cite{width} to the bounds on $\Delta \kappa$. Since the decay
branching ratio of $H \to W^+W^-$ or $ZZ$ for a SM heavy Higgs
boson is large (almost $100\%$ for $m_{H} >
300$\,GeV), the total width measurement can give a strong
constraint on $\kappa$.
We define the accuracy on
the determination of $\kappa$ using the relation
$-n \Delta\Gamma \le \Gamma(\kappa)-\Gamma(\kappa=1)
\le n \Delta\Gamma$,
where $n=1,2$ denoting the
$1\sigma$ and $2\sigma$ accuracy, respectively, and
$\Gamma(\kappa)$ is the width of the Higgs boson for a
general value of $\kappa$, $\Gamma(\kappa=1)$ is for
a SM Higgs boson and $\Delta\Gamma$ is the experimental accuracy
in measuring the Higgs boson width.
From the Table~3 of Ref.~\cite{width}, we find that at the LHC (with an
integrated luminosity of 300~fb$^{-1}$ \cite{LHC-Lumi}), the bounds
on $\Delta \kappa$ obtained from the measurement of the total decay
width of a Higgs boson, via the process $p p \to H \to ZZ \to \mu^+
\mu^- \mu^+ \mu^- $, are as follows:
\begin{eqnarray}                     
&&m_H({\rm GeV})\hspace{1.3cm} 1\sigma \hspace{2.5cm} 2\sigma\nonumber\\
&&
200\--300:~
0.9 \le |\kappa|\le 1.1,\hspace{0.5cm} 0.8 \le |\kappa| \le 1.2  \, .
\,
\label{tevbound'}
\end{eqnarray}

\subsection{Decay Branching Ratio of $H \to \gamma \gamma$}

Another important effect of a non-vanishing anomalous coupling $\kappa$ is
to modify the decay branching ratio of $H \to \gamma \gamma$, and hence,
the production rate of $gg \to H \to \gamma \gamma$ at the
LHC.
In Table~VII, we list the decay
branching ratio $B(H \to \gamma \gamma)$ as a
function of $\Delta \kappa$
for various $m_H$. As shown, $B(H \to \gamma \gamma)$ decreases for
$\Delta \kappa>0$, and increases for $\Delta \kappa < 0$.
For example, for a 120\,GeV Higgs boson, 
 $B(H \to \gamma \gamma)$ decreases by $14\%$ for 
 $\Delta \kappa=0.2$ and increases by $45\%$ for 
 $\Delta \kappa=-0.3$.

\subsection{Detecting a Higgs Boson Resonance}

When the SM Higgs boson is light, it is expected that the production
rate of
$qq \to qqH$ with $H \to W W^* \to \ell \ell' {\not \!\!E_T}$
is large enough to be detected
at the LHC, and this process
can also be used to test the coupling of the $HWW$~\cite{dieter}
by studying the observables near the Higgs boson resonance.
In the effective Lagrangian (\ref{Lagrangian}), the Lorentz structure
(and the dimension) of the anomalous
couplings $\kappa$ and $\kappa'$ is the same as the SM coupling.
Therefore, the result of the study performed in Ref.~\cite{dieter}
also holds for the current study when $m_H$ is less than twice
the $Z$ boson mass.
In this case, its production rate is $\kappa^2 {\cal S}_{SM}
B(\kappa)/B(\kappa=1)$, where ${\cal S}_{SM}$ is the SM rate, and
 $B(\kappa)$ is the decay branching ratio of
$H \to WW^*$ for a general value of $\kappa$.
Needless to say that $B(\kappa=1)$ corresponds to the SM branching ratio.

When the $WW$ (and $ZZ$) decay mode dominates the Higgs boson
decay, such as when $m_H$ is larger than twice $W$ boson
mass, the ratio
of the decay branching ratios for $H \to W W^*$,
$B(\kappa)/B(\kappa=1)$, would be close to 1.
However, this ratio can be quite different from 1 when
$m_H$ is small.
 For example, for $m_H=120$\,GeV,
the ratio $B(\kappa)/B(\kappa=1)$ is 0.68 and 1.4, respectively,
 for $\kappa=0.8$ and $\kappa=1.2$.
From Table~1 of Ref.~\cite{dieter}, one can determine
 the constraint on the ratio $\kappa^2 B(\kappa)/B(\kappa=1)$,
 denoted as $R$, using the relation
${\cal S}_{SM} - n \sqrt{{\cal B}+{\cal S}_{SM}} \le R {\cal S}_{SM}
\le {\cal S}_{SM} + n \sqrt{{\cal B}+{\cal S}_{SM}}$,
where ${\cal B}$ is the background rate and $n=1,2$ denotes the
$1\sigma$ and $2\sigma$ accuracy, respectively.
Assuming an integrated luminosity of 300 fb$^{-1}$, we find that
the bounds on $R$ for various $m_H$ are:
\begin{eqnarray}                  
&&m_H~({\rm GeV})\hspace{1.1cm} 1\sigma \hspace{2.5cm} 2\sigma\nonumber\\
&&110:~
\hspace{0.8cm}
{0.87 \le R}\le 1.13,\hspace{0.3cm} {0.74 \le R}\le
1.26,
\nonumber\\
&&120:~
\hspace{0.8cm}
{0.94 \le R}\le 1.06,\hspace{0.3cm} {0.88 \le R}\le
1.12,
\nonumber\\
&&130:~
\hspace{0.8cm}
{0.97 \le R}\le 1.03,\hspace{0.3cm} {0.93 \le R}\le
1.07,
\nonumber\\
&&150:~
\hspace{0.8cm}
{0.97 \le R}\le 1.03,\hspace{0.3cm} {0.94 \le R}\le
1.06,
\nonumber\\
&&170:~
\hspace{0.8cm}
{0.98 \le R}\le 1.02,\hspace{0.3cm} {0.95 \le R}\le
1.05.
\label{lighthbound}
\end{eqnarray}
To extract the bound on $\kappa$, we need to know the decay
branching ratio $B(\kappa)$ of $ H \to WW^*$ assuming that the
effective Lagrangian (\ref{Lagrangian}) differs from the SM
Lagrangian only in the coefficient
$\kappa$. A few values of $B(\kappa)$ as a function of
$\kappa$ for various $m_H$ are given in Table~VIII.
The result of Tabel~VIII and Eq. (\ref{lighthbound}) indicates that
$\kappa$ can be measured at a few percent level when a
light Higgs boson is detected. However, in our opinion,
this conclusion seems to be too optimistic
because the systematical error of the experiment
at the LHC is expected to be at a similar level of accuracy.

\subsection{At the Linear Collider}

Before closing this section, we note that the anomalous coupling
of $HZZ$ and $HW^+W^-$ 
can also be tested at the LC via the processes 
$e^-e^+ \to Z H(\to b\bar{b})$, 
$e^-e^+ \to e^-e^+  H(\to b\bar{b})$ via $ZZ$ fusion, and 
$e^-e^+ \to \nu {\bar \nu} H(\to b\bar{b})$ via $W^+W^-$ fusion. In
Refs.~\cite{HZZ,BHLMZ}, it was concluded that $\Delta \kappa$ can be
determined better than a percent level for a 120\,GeV SM-like Higgs boson
produced from the above processes, assuming an integrated luminosity 
of $1\, {\rm ab}^{-1}$ for a 500\,GeV (or 800\,GeV) LC.
After converting the notation in Ref.~\cite{BHLMZ} to ours, 
the $2\sigma$ error in measuring $\Delta\kappa$ is 
found to be at the level of $0.3\%$.
Thus the expected precision in the measurement of
$\Delta\kappa$ at a high luminosity LC 
is higher than that at the LHC, in the case of a SM-like Higgs boson 
(i.e., the decay branching ratio of $H \to b {\bar b}$ is about 1).
Even in the case that the Higgs boson is not SM-like, 
the LC can still determine $\Delta \kappa$ by studying the 
Higgs-strahlung process $e^-e^+ \to Z(\to \ell^+ \ell^-) H$ and the
$ZZ$ fusion process $e^-e^+ \to e^-e^+  H$, where $H$ decays into 
anything~\cite{tdr}. 
However, due to its much smaller rate, the sensitivity to 
$\Delta \kappa$ is lower than the SM-like case.  
In the next section,  
we shall show that the precision of determining the dimension-6
anomalous couplings via $VV$ scatterings at the LHC will be 
comparable to the level of the precision expected at the LC.

\section{TESTING ANOMALOUS HVV COUPLINGS\\
         FROM DIMENSION-6 OPERATORS}

\subsection{Anomalous HVV Couplings\\ from Dimension-6 Operators}

Now we consider the test of the anomalous $HVV$ couplings from the
dimension-6 operators arising from the linearly realized effective
Lagrangian. In the linearly realized effective Lagrangian, there
is no gauge invariant anomalous operators at dimension-3 and 
dimension-4, and 
the leading anomalous $HVV$ couplings are from the effective
operators of dimension-6 \cite{linear,G-G}. 
(All the gauge invariant operators with dimension 4 or less have 
been included in the SM Lagrangian.)
In the following, we
shall analyze the test of these leading order anomalous couplings.
As is shown in Refs.\,\cite{linear,G-G}, the {\tt C} and {\tt P}
conserving effective Lagrangian up to dimension-6 operators
containing a Higgs doublet $\Phi$ and the weak bosons $V^a$ is
given by
\begin{equation}                    
{\cal L}_{\mbox{eff}} ~\,=~\, \sum_n \frac{f_n}{\Lambda^2} {\cal O}_n \,,
\label{l:eff}
\end{equation}
where $f_n$'s are the anomalous coupling constants. 
The operators ${\cal O}_n$'s are \cite{linear,G-G}
\begin{eqnarray}                    
&&{\cal O}_{\Phi,1} = \left ( D_\mu \Phi \right)^\dagger
\Phi^\dagger \Phi
\left ( D^\mu \Phi \right ), \nonumber \\
&&{\cal O}_{BW} =  \Phi^{\dagger} \hat{B}_{\mu \nu}
\hat{W}^{\mu \nu} \Phi, \nonumber \\
&& {\cal O}_{DW} = \mbox{Tr}\left (\left[ D_\mu ,\hat{W}_{\nu
\rho} \right]
\left[ D^\mu ,\hat{W}^{\nu\rho} \right]\right), \nonumber \\
&&{\cal O}_{DB}=-\frac{{g^\prime}^2}{2} \left(\partial_\mu
B_{\nu\rho}\right) \left(\partial^\mu
B^{\nu\rho}\right),\nonumber \\
&&{\cal O}_{\Phi,2} =\frac{1}{2}
\partial^\mu\left ( \Phi^\dagger \Phi \right)
\partial_\mu\left ( \Phi^\dagger \Phi \right), \nonumber \\
&&{\cal O}_{\Phi,3} =\frac{1}{3} \left(\Phi^\dagger \Phi
\right)^3,\nonumber\\
&& {\cal O}_{WWW}=\mbox{Tr}[\hat{W}_{\mu
\nu}\hat{W}^{\nu\rho}\hat{W}_{\rho}^{\mu}]
, \nonumber \\
&& {\cal O}_{WW} = \Phi^{\dagger} \hat{W}_{\mu \nu}
\hat{W}^{\mu \nu} \Phi , \nonumber \\
&&{\cal O}_{BB} = \Phi^{\dagger} \hat{B}_{\mu \nu} \hat{B}^{\mu
\nu} \Phi ,  \nonumber \\
&&{\cal O}_W  = (D_{\mu} \Phi)^{\dagger}
\hat{W}^{\mu \nu}  (D_{\nu} \Phi), \nonumber \\
&&{\cal O}_B  =  (D_{\mu} \Phi)^{\dagger} \hat{B}^{\mu \nu}
(D_{\nu} \Phi), \label{O}
\end{eqnarray}                          
where $\hat B_{\mu\nu}$ and $\hat W_{\mu\nu}$ stand for
\begin{eqnarray}
\hat{B}_{\mu \nu} = i \frac{g'}{2} B_{\mu \nu},\;\;\;\;\;\;\;\;
\hat{W}_{\mu \nu} = i \frac{g}{2} \sigma^a W^a_{\mu \nu},
\end{eqnarray}
in which $g$ and $g^\prime$ are the $SU(2)$ and $U(1)$ gauge
coupling constants, respectively.

At tree level, the operators ${\cal O}_{\Phi,1},~{\cal
O}_{BW},~{\cal O}_{DW}$ and ${\cal O}_{DB}$ in Eq.\,(\ref{O})
affect the two-point functions of the weak boson $V$ when $\Phi$
is taken to be its vacuum expectation value:
\begin{eqnarray}                        
\Phi\to \frac{1}{\sqrt{2}}\pmatrix{0\cr v }.
\label{vev}
\end{eqnarray}
Thus they are severely constrained by the precision 
$Z$-pole and low energy data.
In Ref.~\cite{G-G}, it was concluded that at the 
$95\%$ level, in unites of
${\rm TeV}^{-2}$, the magnitude of 
$\frac{f_{DW} }{\Lambda^2}$ or 
$\frac{f_{\Phi,1} }{\Lambda^2}$ is constrained to be 
less than 1, and 
$\frac{f_{DB} }{\Lambda^2}$ or 
$\frac{f_{BW} }{\Lambda^2}$ can be about 
a factor of 10 larger. 
We shall update these bounds in Sec. VIB.
Since it will be very difficult to observe the effect of these 
operators in the high-energy observables, 
in what follows, we will neglect their effect when 
discussing the $VV$ scatterings.

The two operators ${\cal O}_{\Phi,1}$ and ${\cal O}_{\Phi,2}$ in
Eq.\,(\ref{O}) lead to a finite renormalization of the Higgs
boson wave function~\cite{G-G}. Similarly, 
the two operator ${\cal O}_{\Phi,3}$ induces 
a finite renormalization of the Higgs potential~\cite{G-G}. 
In a recent paper~\cite{BHLMZ}, it was shown that after renormalizing  
the Higgs boson field so that the residue of its propagator at its mass
pole is equal to one, two effective anomalous couplings of Higgs boson 
to gauge bosons are induced by the dimension-6 
operator ${\cal O}_{\Phi,2}$ as 
\begin{eqnarray}
{-f_{\Phi,2} (v H +H^2) \over \Lambda^2}
\left[
m_W^2 W^+_\mu W^{-\mu} + \frac{1}{2}m_Z^2 Z_\mu Z^\mu 
\right] \, .
\nonumber
\end{eqnarray}
After converting the above anomalous couplings 
to the notation introduced in Sec.~III for a non-linear 
effective theory, 
the anomalous coupling $\Delta\kappa$ corresponds to 
$\frac{-f_{\Phi,2} v^2 }{ 4 \Lambda^2}$.
For example, at a 500\,GeV (or 800\,GeV) LC with 
an integrated luminosity of $1\, {\rm ab}^{-1}$,
the $2\sigma$ statistical error on the determination of 
the anomalous coupling  
$\frac{f_{\Phi,2} }{\Lambda^2}$ is at 
the level of $0.2 \, {\rm TeV}^{-2}$,
for a 120\,GeV SM-like Higgs boson~\cite{BHLMZ}. 
This corresponds to the determination of 
$\Delta\kappa$ at about the $0.3\%$ level.
Since this operator can only generate $E^2$-dependence 
of the $V_LV_L \to V_LV_L$ scattering amplitude, and it 
is best determined at the LC for a SM-like Higgs boson, 
in what follows, we will neglect its effect in our studies. 

The operator ${\cal O}_{WWW}$ contributes to the  
triple and quartic vector boson self-couplings, but not
the anomalous coupling of Higgs boson to gauge bosons. 
On the other hand, the last
four operators ${\cal O}_{WW},~{\cal
O}_{BB},~{\cal O}_W$ and ${\cal O}_B$ in Eq. (\ref{O}) contribute
to the following anomalous $HVV$ couplings~\cite{G-G}:
\begin{eqnarray}                         
&&{\cal L}^H_{\rm eff}=g_{H\gamma\gamma}HA_{\mu\nu}A^{\mu\nu}
+g^{(1)}_{HZ\gamma}A_{\mu\nu}Z^\mu\partial^\nu H\nonumber\\
&&\hspace{0.4cm}+g^{(2)}_{HZ\gamma}HA_{\mu\nu}Z^{\mu\nu}
+g^{(1)}_{HZZ}Z_{\mu\nu}Z^\mu
\partial^\nu H\nonumber\\
&&\hspace{0.4cm}+g^{(2)}_{HZZ}HZ_{\mu\nu}Z^{\mu\nu}
+g^{(1)}_{HWW}(W^+_{\mu\nu} W^{-\mu}\partial^\nu H+{\rm h.c.})\nonumber\\
&&\hspace{0.4cm}+g^{(2)}_{HWW}HW^+_{\mu\nu}W^{-\mu\nu},
\label{LHeff}
\end{eqnarray}
where
\begin{eqnarray}                           
\displaystyle
&&g^{}_{H\gamma\gamma}=-\bigg(\frac{gm_W}{\Lambda^2}\bigg)\frac{s^2(f_{BB}
+f_{WW})}{2},\nonumber\\
&&g^{(1)}_{HZ\gamma}=\bigg(\frac{gm_W}{\Lambda^2}\bigg)\frac{s(f_W-f_B)}{2c},
\nonumber\\
&&g^{(2)}_{HZ\gamma}=\bigg(\frac{gm_W}{\Lambda^2}\bigg)\frac{s[s^2f_{BB}
-c^2f_{WW}]}{c},\nonumber\\
&&g^{(1)}_{HZZ}=\bigg(\frac{gm_W}{\Lambda^2}\bigg)\frac{c^2f_W+s^2f_B}{2c^2},
\nonumber\\
&&g^{(2)}_{HZZ}=-\bigg(\frac{gm_W}{\Lambda^2}\bigg)\frac{s^4f_{BB}
+c^4f_{WW}}{2c^2},\nonumber\\
&&g^{(1)}_{HWW}=\bigg(\frac{gm_W}{\Lambda^2}\bigg)\frac{f_W}{2},\nonumber\\
&&g^{(2)}_{HWW}=-\bigg(\frac{gm_W}{\Lambda^2}\bigg)f_{WW},
\label{g}
\end{eqnarray}
with $s\equiv \sin\theta_W,~c\equiv \cos\theta_W$.
In our calculation, we have included the complete gauge invariant set of
Feynman diagrams that receive contribution from the anomalous operators
${\cal O}_{WWW},~{\cal O}_{WW},~{\cal O}_{BB},~{\cal O}_W$ and ${\cal O}_B$.
For example, the triple vector boson self-couplings
include the contributions from the operators 
${\cal O}_{WWW}$, ${\cal O}_W$ and ${\cal O}_B$, 
while the quartic vector boson self-couplings are  
induced by ${\cal O}_{WWW}$ and ${\cal O}_W$. 
The reason that the operators ${\cal O}_{WW}$ and ${\cal O}_{BB}$ do not 
modify the gauge boson self-couplings is as follow. 
When the Higgs field of those operators is replaced by its vacuum 
expectation value, it seems that they would induce anomalous operators to 
modify the gauge boson self-couplings. However, the resulting operators are 
proportional to the kinematic term of the $SU(2)$ and $U(1)$ gauge
bosons, and lead to a finite wave-function renormalization of the 
gauge fields by constants 
$Z^{1/2}_{2W}=(1- g^2 f_{WW} v^2 /2 \Lambda^2)^{-1/2}$
and
$Z^{1/2}_{2B}=(1- {g'}^2 f_{BB} v^2 /2 \Lambda^2)^{-1/2}$,
respectively.
Therefore, from the fact that the building blocks of the effective 
Lagrangian ${\cal L}_{\mbox{eff}}$, cf. Eq.~(\ref{l:eff}), 
involving gauge bosons are $g {W}_{\mu \nu}$,
 $g' {B}_{\mu \nu}$, and the covariant derivative $D_\mu$,
 we can perform a finite charge renormalization 
 of the gauge couplings $g$ and $g'$ by constants  
$Z_{g}=(1+ g^2 f_{WW} v^2 /2 \Lambda^2)^{-1/2}$
and
$Z_{g'}=(1+ {g'}^2 f_{BB} v^2 /2 \Lambda^2)^{-1/2}$,
respectively, so that 
the net effect of the operators ${\cal O}_{WW}$ and ${\cal O}_{BB}$ is
to modify only the couplings of a Higgs boson to gauge bosons, but not the 
self-couplings of gauge bosons.
  
In  Eq.\,(\ref{LHeff}), the anomalous $HVV$ couplings are
expressed in terms of the Lorentz-invariant dimension-5 operators
containing the Higgs boson and the gauge bosons $W^\pm,~Z$ and
$\gamma$. Among them, the operators 
$HA_{\mu\nu}A^{\mu\nu}$, $HA_{\mu\nu}Z^{\mu\nu}$, 
$HZ_{\mu\nu}Z^{\mu\nu}$ and $HW^+_{\mu\nu}W^{-\mu\nu}$ 
 can also be induced from the
gauge-invariant dimension-5 operators in the nonlinear realization
of the Higgs sector because in which the Higgs field $H$ is an
electroweak singlet. Thus, it is worth noticing that the following
LHC study of testing the linearly realized anomalous $HVV$
couplings via $VV$ scatterings may be generalized to the case of
the dimension-5 operators in the nonlinear realization.

\subsection{Constraints on ${\mathbf f_n}$
            from the Existing Experiments and the
            Unitarity Requirement}

There are known experiments that can constrain the size of
the anomalous coupling constants $f_n$.

The constraints on the anomalous coupling constants $f_{WWW}$,
$f_{WW}$, $f_{BB}$, $f_W$ and $f_B$ have been studied in
Refs.\cite{G-G,hisz,ADS}. At the tree level, $\Delta S$ and $\Delta T$
are proportional to $f_{BW}/\Lambda^2$ and $f_{\Phi,1}/\Lambda^2$,
respectively. Thus we can obtain the $68\%$ and $95\%$  bounds
on the ($f_{BW}/\Lambda^2$)-($f_{\Phi,1}/\Lambda^2$) plane
directly from the corresponding bounds in Fig. 2(a). This is shown
in Fig. 4. We see from Fig. 4 that the precision data give quite
strong constraints on $f_{BW}/\Lambda^2$ and
$f_{\Phi,1}/\Lambda^2$. At the one loop level, $\Delta S$ and $\Delta
T$ are related to other five anomalous coupling constants through
loop corrections. Following Refs.\cite{G-G,hisz,ADS}, we make a
one parameter fit of the five anomalous coupling constants by
using the formulas given in Ref.\cite{ADS} and the updated $\Delta
S$-$\Delta T$ bounds in Fig. 2(a). The obtained $95\%$ C.L.
constraints (in units of TeV$^{-2}$) for $m_H=100$ GeV are
\begin{eqnarray}                              
&& -6\leq f_{WWW}/\Lambda^2\leq 3, \nonumber\\
&& -6\leq f_{W}/\Lambda^2\leq 5,  \nonumber\\
&& -4.2\leq f_{B}/\Lambda^2\leq 2.0,  \nonumber\\
&& -5.0\leq f_{WW}/\Lambda^2\leq 5.6,  \nonumber\\
&& -17\leq f_{BB}/\Lambda^2\leq 20. \label{EWconstraints}
\label{LEPbounds}
\end{eqnarray}
These constraints are much weaker than that shown in Fig. 4
because these five anomalous coupling constants contribute to
$\Delta S$ and $\Delta T$ through one loop corrections which are
suppressed by the loop factor $1/16\pi^2$.

The triple gauge coupling data lead to the following $95\%$ C.L.
constraints (in units of
TeV$^{-2}$)\cite{HgammagammaLHC,G-G,wwv:teva,wwv:lep}:
\begin{eqnarray}                             
&&-31\leq (f_W+f_B)/\Lambda^2\leq 68\;  \; \mbox{ for $f_{WWW}
=0$},
 \nonumber \\
&& -41 \leq f_{WWW}/\Lambda^2\leq 26 \; \; \mbox{  for
$f_W+f_B=0$}.
\label{limitf:wwv}
\end{eqnarray}

Furthermore, the Higgs searches data at the LEP2 and the Tevatron
can give rise to the following $95\%$ C.L. bound on $f_{WW(BB)}$ (in
units of TeV$^{-2}$) for $m_H\le 150$ GeV \cite{HgammagammaLHC}
\begin{eqnarray}                            
-7.5\le \frac{f_{WW(BB)}}{\Lambda^2}\le 18.
\label{Higgssearchbound}
\end{eqnarray}

The theoretical constraint on $f_n$ coming from the requirement of
the unitarity of the $S$ matrix has been studied in
Ref.\cite{unitarity}. In terms of the present symbols of the
anomalous coupling constants, the unitarity bounds given in
Ref.\cite{unitarity} read (in units of TeV$^{-2}$)
\begin{eqnarray}                                     
\displaystyle \left|\frac{f_B}{\Lambda^2}\right|\le
\frac{98}{\Lambda^2},\;\;\;\;\;\;\;\;\;\;\;
\left|\frac{f_W}{\Lambda^2}\right|\le\frac{31}{\Lambda^2},
\nonumber
\end{eqnarray}
\begin{eqnarray}
-\frac{784}{\Lambda^2}+\frac{3556m_W}{\Lambda^3}\le
\frac{f_{BB}}{\Lambda^2}\le
\frac{638}{\Lambda^2}+\frac{3733m_W}{\Lambda^3},\nonumber
\end{eqnarray}
\begin{eqnarray}
\left|\frac{f_{WW}}{\Lambda^2}\right|\le
\frac{35.2}{\Lambda^2}+\frac{4.86}{\Lambda m_W},\;\;\;\;\;\;
\left|\frac{f_{WWW}}{\Lambda^2}\right|\le\frac{38}{3g^2\Lambda^2},
\label{Uconstraint}
\end{eqnarray}
in which we have put the center-of-mass energy $\sqrt{s}\approx
\Lambda$. For $\Lambda\approx 2$ TeV, the bounds are (in units of
TeV$^{-2}$)
\begin{eqnarray}                      
\displaystyle \left|\frac{f_B}{\Lambda^2}\right|\le
24.5\;\;\;\;\left|\frac{f_W}{\Lambda^2}\right|\le
7.8\;\;\;\;\left|\frac{f_{WWW}}{\Lambda^2}\right|\le 7.5,\nonumber
\end{eqnarray}
\begin{eqnarray}
\displaystyle -160\le \frac{f_{BB}}{\Lambda^2}\le
197,\;\;\;\;\;\;\left|\frac{f_{WW}}{\Lambda^2}\right|\le 39.2~.
\label{Ubounds}
\end{eqnarray}
These bounds are essentially of the same level as the above bounds
(\ref{EWconstraints}) and (\ref{limitf:wwv}).

We see that, except for the constraints from the precision data
(cf. Fig. 4), all other constraints are rather weak. Furthermore,
the above constraints on $f_n/\Lambda^2$ lead to the following
constraints (in units of TeV$^{-1}$) on the anomalous coupling
constants $g^{(1)}_{HWW},~g^{(2)}_{HWW},~g^{(1)}_{HZZ}$ and
$g^{(2)}_{HZZ}$, cf. Eq.~(\ref{g}) , related to the $VV$ scatterings:
\begin{eqnarray}                                
&&-0.20\le g^{(1)}_{HWW}\le 0.065,\nonumber\\
&&-1.2\le -g^{(2)}_{HWW}\le 0.73,\nonumber\\
&&-0.26\le g^{(1)}_{HZZ}\le 0.24,\nonumber\\
&&-0.62\le -g^{(2)}_{HZZ}\le 0.36.
\label{constraint-g}
\end{eqnarray}
We shall see in the following subsection that the limits on these
coupling constants obtained from studying the 
$VV$ scattering processes at the LHC will be significantly 
stronger than those in Eq. (\ref{constraint-g}).

\subsection{Testing ${\mathbf f_n}$ via VV Scatterings at the LHC}

The test of the anomalous $HVV$ couplings from the dimension-6
operators via $VV$ scatterings is quite different from that for
the dimension-3 operator. The relevant operators in
Eq.\,(\ref{O}) contain two derivatives, so that the interaction
vertices 
themselves behave as $E^2$ at high energies. Thus, at high
energies, the scattering amplitudes of $V_LV_L \to V_LV_L$
grows as $E^4$, and those containing transverse components,
i.e., $V_TV_T \to V_LV_L$, $V_TV_L \to V_TV_L$ and $V_LV_L \to V_TV_T$,
grow as $E^2$. Hence the scattering processes containing
$V_T$ actually behave as {\it signals} rather than backgrounds.
This is very different from the case of the nonlinearly 
realized dimension-3 anomalous
coupling studied in Sec. IV. As discussed in Sec. II, since the
$V_TV_T$ and $V_LV_T$ scatterings also contribute to the signal
rate, we decide to do the full $2 \to 4$ tree level calculation,
in contrast to performing the calculation using the EWA folded by
the $2 \to 2$ $VV$ scattering amplitudes. Nevertheless, we shall
use the EWA to check the high energy behavior of the full
calculation, in the case that the anomalous coupling is large.

We calculate the tree level cross sections for all the processes
listed in Eq. (\ref{channels}) with the same method described in
Sec. IV, but for the linearly realized effective Lagrangian
theory. Since there are several dimension-6 anomalous couplings related
to the $VV$ scatterings in this theory [cf. Eq. (\ref{O})], the
analysis is more complicated than in the 
nonlinear realization case. If the
anomalous coupling constants are of the same order of magnitude,
the interferences between them may be significant, depending on
the relative phases among them. This undoubtedly complicates the
analysis. In the following, we again perform the single parameter analysis,
i.e., assuming only one of the anomalous coupling constants is
dominant at a time. We see from Eq.~(\ref{O}) that the coupling
constants related to $VV$ scatterings are $f_{WWW}$, $f_{WW}$,
$f_{BB}$, $f_{W}$ and $f_{B}$. 
As discussed in Sec.~VIA, the operator 
${\cal O}_{WWW}$ does not induce the anomalous coupling of a Higgs boson to 
gauge bosons, and it may not be directly related to the electroweak 
symmetry breaking mechanism, 
so we assume it is small in the following analysis. 
Detailed calculations show that the
contributions of $f_B$ and $f_{BB}$ to the most sensitive $pp\to
W^+W^+jj\to l^+\nu l^+\nu jj$ channel are small even if they
are of the same order of magnitude as the anomalous coupling
constants $f_W$ and $f_{WW}$. 
Hence, we shall ignore their contributions in the following analysis, 
and discuss only the sensitivity of 
the $VV$ scatterings to the measurement of 
$f_W$ and $f_{WW}$.

In the case that $f_W$ dominates, the obtained numbers of events
at the LHC with an integrated luminosity of $300$ fb$^{-1}$,  
for various values
of $m_H$ and $f_W/\Lambda^2$, are listed in Table IX(A) and Tables
X$-$XII. From these Tables, we see that the most sensitive
channel is still $pp\to W^+W^+jj\to l^+\nu l^+\nu jj$, as
listed in Table IX(A). All other channels are less interesting.
Similar to the numbers in Table III, the number of the SM events 
(for $f_W/\Lambda^2=0$) is taken as the 
intrinsic electroweak backgrounds, i.e., $N_B=N(f_W=0)$. The
number of the signal events (for $f_W/\Lambda^2 \neq 0$)
is then defined as $N_S=N(f_W\ne 0)-N_B$.
We also list the values of the deviation of the signal from the
total statistical fluctuation, $N_S/\sqrt{N_S+N_N}$, in the
parentheses in Table IX(A). If no anomalous coupling effect is found
via this process, we can set the following bounds on 
$f_W/\Lambda^2$ (in units of TeV$^{-2}$):
\begin{eqnarray}                       
&&1\sigma:~~~~~~-1.0< f_W/\Lambda^2< 0.85 \, , \nonumber\\
&&2\sigma:~~~~~~-1.4< f_W/\Lambda^2\leq 1.2, \label{fW}
\end{eqnarray}
for 115 GeV$\alt m_H\alt$
300 GeV.

In the case that $f_{WW}$ dominates, the numbers of
events are listed in Table IX(B), and the corresponding bounds are (in
units of TeV$^{-2}$) as follows:
\begin{eqnarray}                       
&&1\sigma:~~~~~~-1.6\leq f_{WW}/\Lambda^2<1.6,\nonumber\\
&&2\sigma:~~~~~~-2.2\leq f_{WW}/\Lambda^2< 2.2, \label{fWW}
\end{eqnarray}
which are somewhat weaker than those in Eq.~(\ref{fW}).

To see the effect of interference, we take a special case as an
example, in which $f_W=-f_{WW}\equiv f$, i.e., $g^{(1)}_{HVV}$ and
$g^{(2)}_{HVV}$ in Eq. (\ref{g}) are of the same sign. Then we
obtain the following bounds (in units of TeV$^{-2}$):
\begin{eqnarray}                       
&&1\sigma:~~~~~~-0.6\leq f/\Lambda^2\leq
0.5,\nonumber\\
&&2\sigma:~~~~~~-0.9\leq f/\Lambda^2\leq 0.75.
\label{f}
\end{eqnarray}
In this case, the interference enhances the sensitivity. Of
course, if $f_W=f_{WW}=f$, the sensitivity will be reduced.

From the bounds in Eqs. (\ref{fW}), (\ref{fWW}), together with
the relations in Eq. (\ref{g}), we obtain the corresponding
bounds on $g^{(i)}_{HVV},~i=1,2$ (in units of TeV$^{-1}$):
\begin{eqnarray}                           
&&1\sigma:\nonumber\\
&&\hspace{0.4cm}-0.026< g^{(1)}_{HWW}< 0.022,\nonumber\\
&&\hspace{0.4cm}-0.026< g^{(1)}_{HZZ}< 0.022,\nonumber\\
&&\hspace{0.4cm}-0.014< g^{(1)}_{HZ\gamma}< 0.012,\nonumber\\
&&\hspace{0.4cm}-0.083\leq g^{(2)}_{HWW}< 0.083,\nonumber\\
&&\hspace{0.4cm}-0.032\leq g^{(2)}_{HZZ}< 0.032,\nonumber\\
&&\hspace{0.4cm}-0.018\leq g^{(2)}_{HZ\gamma}< 0.018,\nonumber\\
&&2\sigma:\nonumber\\
&&\hspace{0.4cm}-0.036< g^{(1)}_{HWW}\leq 0.031,\nonumber\\
&&\hspace{0.4cm}-0.036< g^{(1)}_{HZZ}\leq 0.031,\nonumber\\
&&\hspace{0.4cm}-0.020< g^{(1)}_{HZ\gamma}\leq 0.017,\nonumber\\
&&\hspace{0.4cm}-0.11~\leq g^{(2)}_{HWW}< ~0.11,\nonumber\\
&&\hspace{0.4cm}-0.044\leq g^{(2)}_{HZZ}< 0.044,\nonumber\\
&&\hspace{0.4cm}-0.024\leq g^{(2)}_{HZ\gamma}< 0.024.
\label{gbounds}
\end{eqnarray}
These bounds are to be compared with the $1\sigma$ bound on
$g^{(2)}_{HWW}$ obtained from studying the on-shell Higgs boson production
via weak boson fusion at the LHC, as given recently in
Ref.\cite{PRZ}. In Ref.\cite{PRZ}, $g^{(2)}_{HWW}$ is
parametrized as $g^{(2)}_{HWW}$$=1/\Lambda_{5}=g^2
v/\Lambda^2_{6}$, and the obtained $1\sigma$ bound on $\Lambda_6$
for an integrated luminosity of 100 fb$^{-1}$ is about $\Lambda_6
\agt$ 1 TeV, which corresponds to $g^{(2)}_{HWW}=1/\Lambda_5\le
0.1~ {\rm TeV}^{-1}$. We see that the $1\sigma$ bounds listed in
Eq. (\ref{gbounds}) are all stronger than this bound. For an
integrated luminosity of 300 fb$^{-1}$, the bound
$g^{(2)}_{HWW}=1/\Lambda_5\le 0.1~ {\rm TeV}^{-1}$ given in
Ref.\cite{PRZ} corresponds roughly to a 1.7$\sigma$ level of 
accuracy. Comparing it with the results in Eq.
(\ref{gbounds}), we conclude that our 2$\sigma$ bound on 
$g^{(2)}_{HWW}$ is at about the same level of accuracy, 
while our 2$\sigma$ bounds on
the other five $g^{(i)}_{HVV}~(i=1,2)$ are all stronger than 
those given in Ref.\cite{PRZ}.

It has been shown in Ref.\cite{HZZ} that the anomalous $HZZ$
coupling constants $g^{(1)}_{HZZ}$ and $g^{(2)}_{HZZ}$ can be
tested rather sensitively at the LC via the Higgs-strahlung
process $e^+e^-\to Z^*\to Z+H$ with $Z\to f\bar{f}$.
In Ref.\cite{HZZ}, $g^{(1)}_{HZZ}$ and $g^{(2)}_{HZZ}$ are
parametrized as $g^{(1)}_{HZZ}=\displaystyle\frac{g_Z}{m_Z}c_V$
and $g^{(2)}_{HZZ}=\displaystyle\frac{g_Z}{m_Z}b_Z$, respectively,
The obtained limits on the coefficients $c_V$ and $b_V$ are
$c_V\sim b_V\sim O(10^{-3})$ \cite{HZZ} which correspond to the
limits $g^{(1)}_{HZZ}\sim g^{(2)}_{HZZ}\sim O(10^{-3}\--10^{-2})$
TeV$^{-1}$. Although the LHC $W^+W^+$ scattering bounds shown in
Eq. (\ref{gbounds}) are weaker than these LC bounds, $W^+W^+$
scattering at the LHC can provide the bounds on
$g^{(i)}_{HWW},~i=1,2$. So that the two experiments are
complementary to each other.

To verify that our calculation does give the correct asymptotic
behavior as that given in Ref.\cite{unitarity}, and to see the
difference between the complete tree level result and the EWA
result (with only the $V_LV_L \to V_LV_L$ contribution included),
we plot in Fig. \ref{MWW} the $M_{WW}$ distributions in the
$W^+W^+$ channel with $f_W/\Lambda^2=5$ TeV$^{-2}$ from the
complete tree level calculation (solid curve), 
the EWA calculation with the exact $W^+_LW^+_L \to W^+_LW^+_L$
amplitude (dashed curve), and the result from the EWA
calculation with the asymptotic formula \cite{unitarity} for
the $W^+_LW^+_L \to W^+_LW^+_L$ scattering 
amplitude\footnote{ It only contains
terms proportional to $E^4$ or $E^2$. The $E^0$ contribution is
not included.} (dotted curve). We see that the three curves
coincide at high energies which indicates that the asymptotic
behavior of the complete tree level result obtained numerically is
correct. The dashed curve is significantly below the solid curve at
lower energies, which shows that the signal from the transverse
component contributions taken into account in the complete tree
level calculation is very important. The dotted curve is much lower
than the dashed curve at low energies even though they are all from the
EWA approach with only the longitudinal component contributions
taken into account. This is because that there are contributions
of energy-independent terms contained in the dashed curve which are
not included in the dotted curve, and the energy-independent terms
cause the dashed curve to 
peak significantly in the low energy region due to the larger
parton luminosities in the smaller $M_{WW}$ region. To check the
correctness of this explanation, we have calculated not only the
$M_{WW}$ distribution with a constant amplitude which shows the
above-mentioned peak, but also the $M_{WW}$ distributions for the two
EWA curves with a very large $f_W/\Lambda^2$, say
$f_W/\Lambda^2=100$ TeV$^{-2}$, with which the energy-independent
terms are unimportant. Indeed, the two obtained curves in this
case almost completely coincide, and the peak is shifted
significantly to the high energy region. Finally, we note that for
the values of the anomalous couplings relevant to the bounds given
in Eqs. (\ref{fW}) and (\ref{fWW}), the unitarity condition is
well respected, so that our full $2\to 4$ calculation is
justified. Furthermore, since in our full $2 \to 4$ calculation,
we keep track on the polarization states of the final state $V$
bosons, we can also predict the kinematical distributions of the
final state leptons. As an example, Fig.\,6 shows the
distribution of the invariant mass of the dileptons from the
decay of the final state $W^+$ bosons produced via $pp \to
W^+W^+jj$ for various scenarios of the anomalous couplings.
It indicates that 
examining in detail various kinematical distributions might help
to distinguish various scenarios of the new physics effect.

\section{CONCLUSIONS}

We have examined the possibility of testing the anomalous $HVV$
couplings of a light Higgs boson with mass $115~{\rm GeV}\le
m_H\le 300~{\rm GeV}$ at the LHC via various channels of $VV$
scatterings. This type of test is of special interest if the
anomalous $HVV$ couplings differ from that of the SM by a
significant amount so that the direct detection of the Higgs
resonance is difficult. The study includes two types of anomalous
$HVV$ couplings, namely the anomalous $HVV$ couplings from the
nonlinear dimension-3 operator \cite{CK} and anomalous $HVV$
couplings from the linearly realized dimension-6
operators\,\cite{linear,G-G}. The gold-plated pure leptonic modes
are chosen for detecting the final state weak bosons. To reduce
various kinds of backgrounds, we impose the kinematical cuts
suggested in Ref.\,\cite{WW95}.~\footnote{Although it is
possible to refine the kinematic cuts to further enhance the ratio
of signal to background rates, for simplicity, we applied exactly
the same kinematic cuts as those proposed in Ref.\cite{WW95} in
this study.} The calculations are carried out numerically for the
full cross sections of $pp\to VVjj$ including both the signals and
the backgrounds under the kinematical cuts, and we take into
account only the statistical uncertainties in this calculation.
The results show that, with a sufficiently high integrated
luminosity, such as 300 fb$^{-1}$,
the tests of the anomalous $HVV$ couplings can be rather
sensitive. We note that to further discriminate the effect of the
anomalous $HVV$ coupling from that of a strongly interacting EWSB
sector with no light resonance will eventually demand a
multichannel analysis at the LHC by searching for the light Higgs
resonance through all possible on-shell production channels
including gluon-gluon fusion. Once the light Higgs resonance is
confirmed, $VV$ scatterings, especially the $W^+W^+$ channel, can
provide rather sensitive tests of various anomalous $HVV$
couplings for probing the EWSB mechanism.

In the nonlinearly realized Higgs sector\,\cite{CK}, the leading
anomalous $HVV$ coupling ($\kappa$) comes from the dimension-3 operator
$\frac{1}{2}\kappa vHD_\mu\Sigma^\dagger D^\mu\Sigma$.
The difference $\Delta\kappa\equiv
\kappa-1$ represents the deviation from the SM value
$\kappa=1$.
Our calculation shows that the most
sensitive channel for testing this anomalous coupling is $pp\to
W^+W^+jj\to \l^+\nu \l^+\nu jj$.  We see from Table\,III that
 the LHC can constrain $\Delta\kappa$ to the range $~-0.3\le
\Delta\kappa\le 0.2~$ in the case that
 no anomalous coupling effect is detected
in the channel $pp\to \l^+\nu \l^+\nu jj$. For comparison, several
possible tests of $\Delta\kappa$ in the high energy collider
experiments are also discussed in Sec.\,III. A summary of all the
constraints considered in Sec.\,III is listed in Table\,XIII.

In the linearly realized Higgs sector\,\cite{G-G}, the leading
anomalous $HVV$ couplings arise from a set of dimension-6
operators, $\frac{f_W}{\Lambda^2}{\cal O}_W$,
~$\frac{f_{WW}}{\Lambda^2}{\cal O}_{WW},~$
$\frac{f_B}{\Lambda^2}{\cal O}_B$, and
$~\frac{f_{BB}}{\Lambda^2}{\cal O}_{BB}$, as shown in
Eq.\,(\ref{O}). The anomalous $HWW$ and $HZZ$ couplings are
expressed in Eq.\,(\ref{LHeff}) in terms of the anomalous
coupling constants $g^{(1)}_{HWW}$, $g^{(2)}_{HWW}$,
$g^{(1)}_{HZZ}$ and $g^{(2)}_{HZZ}$ which are connected to the
above anomalous coupling constants $f_W$, $f_{WW}$, $f_{WWW}$,
$f_B$, and $f_{BB}$ via Eq.\,(\ref{g}) . These anomalous coupling
constants are constrained by the precision electroweak data, the
triple gauge coupling data, and the requirement of the unitarity
of the $S$ matrix. Such constraints on the coupling constants
$f_n$ are given in Eqs.\,(\ref{LEPbounds})--(\ref{Ubounds}). The
corresponding constraints on $g^{(1)}_{HWW}$, $g^{(2)}_{HWW}$,
$g^{(1)}_{HZZ}$, and $g^{(2)}_{HZZ}$ are shown in
Eq.\,(\ref{constraint-g}) which restricts these anomalous
couplings to be of $O(10^{-1})-O(1)$. Our calculation shows that
the anomalous $HWW$ couplings can be sensitively tested via $pp\to
W^+W^+ jj\to \l^+\nu \l^+\nu jj$. Since there are several
anomalous couplings from the dimension-6 operators contributing to the
$VV$ scatterings, the analysis of the LHC sensitivity is more
complicated than in the nonlinearly realized case. If the
anomalous couplings are of the same order of magnitude, the
interferences depending on their relative signs could be
significant. In this study, we made a single parameter analysis,
i.e., assuming only one of the anomalous coupling constants is
dominant at a time. Detailed calculations showed that the
contributions of $f_B$ and $f_{BB}$ to the most sensitive $pp\to
W^+W^+jj\to \l^+\nu \l^+\nu jj$ channel are very small even
if they are of the same order of magnitude as other anomalous
coupling constants. So, $f_B$ and $f_{BB}$ are not particularly 
useful in our
analysis. We then analyzed the two parameters $f_W$ and $f_{WW}$,
separately. The $1\sigma$ and $2\sigma$ bounds on these two
anomalous couplings via the most sensitive channel $pp\to
W^+W^+jj\to \l^+\nu \l^+\nu jj$ are listed in
Eqs.\,(\ref{fW}) and (\ref{fWW}), and the corresponding bounds on
the anomalous couplings $g^{(i)}_{HVV},~i=1,2$ are given in
Eq.\,(\ref{gbounds}), i.e., the $1\sigma$ bounds are of
$O(10^{-2})$, and the $2\sigma$ bounds are of
$O(10^{-2}\--10^{-1})$. These bounds are stronger than that
obtained from studying the on-shell 
Higgs boson production via weak boson fusion given in
Ref.\,\cite{PRZ}. They are also complementary to the bounds on
$g^{(1)}_{HZZ}$ and $g^{(2)}_{HZZ}$ at the Linear Collider (LC),
as given in Ref.\,\cite{HZZ}. A summary of all the constraints
considered in Sec.\,IV is displayed in Table\,XIV.

In summary, we find that  $VV$ scatterings are not only important
for probing the strongly interacting electroweak symmetry breaking
mechanism when there is no light Higgs boson, but also valuable for
sensitively testing of the anomalous $HVV$ couplings (especially
anomalous $HWW$ coupling) at the LHC when there is a
light Higgs boson in the mass range of $115-300$\,GeV.
This further supports the ``no-lose'' theorem\,\cite{mike} for the
LHC to decisively probe the electroweak symmetry breaking mechanism.

\vspace*{5mm}
\noindent
{\bf Acknowledgements}~~
We would like to thank Jens Erler and Peter B. Renton
for discussing the electroweak precision data,
Tao Han for interesting discussions and for
providing related calculation codes; and Dieter Zeppenfeld
for discussing the weak boson physics.
B.Z. and Y.P.K. are supported
by the National Natural Science Foundation of China
(under Grant 90103008) and
Foundation of Fundamental Research of Tsinghua University; H.J.H. and C.P.Y.
are supported by the
Department of Energy of the USA under Grant DEFG0393ER40757, and
the National Science Foundation of USA under Grant PHY-0100677,
respectively.

\onecolumn

\section*{TABLES}

\begin{table}[H]
\null\noindent{\small Table I. The $95\%$\,C.L. 
limits on $\Delta\kappa$ imposed by $Z$-pole and low energy precision
electroweak data, for a few 
typical values of new physics scale  $\Lambda$ and Higgs 
boson mass $m_H^{~}$.
}
\vspace*{3mm}
\begin{center}
\begin{tabular}{cccc}
&&&\\[-2.5mm]
$m_H$(GeV) & &$\Lambda$(TeV)  & \\
 &$1$  & $10$   &  $100$\\ [1.5mm]
\hline 
&&&\\[-2.5mm]
115& ~$-0.15\leq \Delta\kappa \leq 0.23$~
& ~$-0.069\leq \Delta\kappa \leq 0.12$~
& ~$-0.045\leq \Delta\kappa \leq 0.08$~\,
\\ 
300& $0.074\leq \Delta\kappa \leq 0.60$
& $ 0.027\leq \Delta\kappa \leq 0.24$
& $ 0.016\leq \Delta\kappa \leq 0.15$
\\ 
800& (excluded)
& $ 0.20\leq \Delta\kappa \leq 0.45$
& $ 0.11\leq \Delta\kappa \leq 0.26$
\\[1.5mm]
\end{tabular}
\end{center}
\end{table}

\begin{table}[H]
\null\noindent{\small Table II. Cross section 
$\sigma_{\lambda_1 \lambda_2}$ (in units of fb) 
of $pp \to W^+_{\lambda_1} W^+_{\lambda 2} jj$ at the LHC 
for various values of $\Delta\kappa$ with $m_H=115$ GeV.
$W^+_\lambda$ denotes a polarized $W$ boson with 
the polarization index $\lambda=T$ or $L$.}
\vspace*{3mm}
\begin{center}
\begin{tabular}{ccccc}
&&&\\[-2.5mm]
$\Delta\kappa$&$\sigma_{\rm all}$&$\sigma_{LL}$&$\sigma_{LT}$&$\sigma_{TT}$\\
[1.5mm]
\hline 
&&&\\[-2.5mm]
0.0&1.1&0.02&0.2&0.8\\ 
0.4& 4.1&3.0&0.3&0.8\\[1.5mm]
\end{tabular}
\end{center}
\end{table}

\begin{table}
 \tabcolsep 3pt
\null\noindent
{\small Table III. Number of events at the LHC, 
with an integrated luminosity of
$300~{\rm fb}^{-1}$, for $pp\to W^{\pm} W^{\pm} jj$
$\to l^\pm \nu(\bar{\nu})l^\pm \nu(\bar{\nu}) jj$ ($l^\pm=e^\pm$ or
$\mu^\pm$) with various values of $m_H$ and $\Delta\kappa$.
($\Delta\kappa=0$ corresponds to the SM.) The values of
$N_S$/$\sqrt{N_S+N_B}$ are also shown in the parentheses.}
\vspace*{4mm} {\small
\begin{tabular}{ccccccccccc}
$pp\to W^+W^+jj\to l^+\nu l^+\nu jj$& & & & & & & & & &\\
\hline\hline
$m_H$(GeV)& & & & & $\Delta\kappa$& & & & &\\
 &-1.0&-0.6&-0.3&0.0&0.2&0.3&0.4&0.5&0.6&0.7~~~~~~\\
           & & & & & & & & & &~~~~~~\\[-2.5mm]
\hline
& & & & & & & & & &~~~~~~\\[-2.5mm]
 115 & 62 (6.0) & 48 (4.8)& 27 (2.3)& 15& 24 (1.8)& 37 (3.6)& 58 (5.6)
 &$-$&$-$&$-$~~~~~~\\
 130&62 (6.0)&48 (4.8)&27 (2.3)&15&24 (1.8)&37 (3.6)&57 (5.5)
 &$-$&$-$&$-$~~~~~~\\
 200&62 (6.0)&48 (4.8)&28 (2.5)&15&22 (1.5)&33 (3.1)&52 (5.1)&78 (7.1)
 &$-$&$-$~~~~~~\\
 300&61 (5.9)&49 (4.9)&30 (2.7)&16&20 (1.1)&29 (2.6)&43 (4.3)&65 (6.2)
 &95 (8.2)&136 (10.4)~~~~~~\\[2mm]
\hline\hline
$pp\to W^-W^- jj\to l^-\bar{\nu}l^-\bar{\nu}jj$& & & & & & & & & &\\
\hline\hline
$m_H$(GeV)& & & & &$\Delta\kappa$ & & & & &\\
 &-1.0&-0.6&-0.3&0.0&0.2&0.3&0.4&0.5&0.6&0.7~~~~~~\\
           & & & & & & & & & &~~~~~~\\[-2.5mm]
\hline
& & & & & & & & & &~~~~~~\\[-2.5mm]
 115 & 13 & 10 & 6 & 3 & 4 & 6 & 9 &$-$&$-$&$-$~~~~~~\\
 130 & 13 & 10 & 6 & 3 & 4 & 6 & 9 &$-$&$-$&$-$~~~~~~\\
 200 & 13 & 10 & 6 & 3 & 4 & 6 & 8 & 11 &$-$&$-$~~~~~~\\
 300 & 13 & 10 & 6 & 4 & 4 & 5 & 7 & 10 & 14 & 19~~~~~~\\[2mm]
\end{tabular}}
\end{table}

\begin{table}
\noindent {\small Table IV. Number of events at the LHC, with an
integrated luminosity of 300 fb$^{-1}$, for $pp\to
W^+W^-jj\to l^+\nu l^-\nu jj$ ($l^\pm=e^\pm$ or $\mu^\pm$) with
various values of $m_H$ and $\Delta\kappa$.
($\Delta\kappa=0$ corresponds to the SM.)} 
\vspace*{0.4cm}
\begin{tabular}{ccccccccccc}
$m_H$(GeV)& & & & &$\Delta\kappa$ & & & & &\\
 &-1.0&-0.6&-0.3&0.0&0.2&0.3&0.4&0.5&0.6&0.7~~~~~~\\
           & & & & & & & & & &~~~~~~\\[-2.5mm]
\hline
& & & & & & & & & &~~~~~~\\[-2.5mm]
 115 & 19 & 14 & 8 & 4 & 7 & 11 & 17 &$-$&$-$
 &$-$~~~~~~\\
 130 & 19 & 14 & 8 & 4 & 7 & 11 & 17 &$-$&$-$
 &$-$~~~~~~\\
 200 & 19 & 14 & 8 & 4 & 8 & 12 & 19 & 29
 &$-$&$-$~~~~~~\\
 300 & 19 & 14 & 7& 4 & 9 & 14 & 23 & 34
 & 48 & 69~~~~~~\\[2mm]
\end{tabular}
\end{table}

\begin{table}
\null\noindent {\small Table V. Number of events at the LHC, with
an integrated luminosity of 300 fb$^{-1}$, for $pp\to
ZW^\pm jj\to l^+l^-l^\pm\nu (\bar{\nu}) jj$ ($l^\pm=e^\pm$ or
$\mu^\pm$) with various values of $m_H$ (in GeV) and
$\Delta\kappa$.
($\Delta\kappa=0$ corresponds to the SM.)}
\vspace*{0.4cm}
\begin{tabular}{ccccccccccc}
$pp\to ZW^+ jj\to l^+l^-l^+\nu jj$& & & & & & & &\\
\hline\hline
$m_H$(GeV)& & & & &$\Delta\kappa$ & & & & &\\
 &-1.0&-0.6&-0.3&0.0&0.2&0.3&0.4&0.5&0.6&0.7~~~~~~\\
           & & & & & & & & & &~~~~~~\\[-2.5mm]
\hline
& & & & & & & & & &~~~~~~\\[-2.5mm]
 115 & 9 & 7 & 4 & 2 & 3 & 5 & 7 &$-$&$-$&$-$~~~~~~\\
 130 & 9 & 7 & 4 & 2 & 3 & 5 & 7 &$-$&$-$&$-$~~~~~~\\
 200 & 9 & 7 & 4 & 2 & 3 & 4 & 6 & 10 &$-$&$-$~~~~~~\\
 300 & 9 & 7 & 4 & 2 & 3 & 4 & 5 & 9 & 12 & 16~~~~~~\\[2mm]
\hline\hline
$pp\to ZW^- jj\to l^+l^-l^-\bar{\nu} jj$& & & & & & & &\\
\hline\hline
$m_H$(GeV)& & & & &$\Delta\kappa$ & & & & &\\
&-1.0&-0.6&-0.3&1.0&0.2&0.3&0.4&0.5&0.6&0.7~~~~~~\\
           & & & & & & & & & &~~~~~~\\[-2.5mm]
\hline
& & & & & & & & & &~~~~~~\\[-2.5mm]
 115 & 4 & 3 & 2 & 1 & 1 & 2 & 3 &$-$&$-$&$-$~~~~~~\\
 130 & 4 & 3 & 2 & 1 & 1 & 2 & 3 &$-$&$-$&$-$~~~~~~\\
 200 & 4 & 3 & 2 & 1 & 1 & 2 & 3 & 4 &$-$&$-$~~~~~~\\
 300 & 4 & 3 & 2 & 1 & 1 & 2 & 2 & 4 & 5 & 7~~~~~~\\[2mm]
\end{tabular}
\end{table}

\begin{table}
\null\noindent {\small Table VI. Number of events at the LHC, with
an integrated luminosity of 300 fb$^{-1}$ for $pp\to
ZZjj\to l^+l^-l^+l^-(\nu\bar{\nu}) jj$ ($l^\pm=e^\pm$ or
$\mu^\pm$) with various values of $m_H$ (in GeV) and
$\Delta\kappa$. ($\Delta\kappa=0$
corresponds to the SM.)} \vspace*{0.4cm}
\begin{tabular}{ccccccccccc}
$pp\to ZZjj\to l^+l^-l^+l^- jj$ & & & & & & & & &\\
\hline\hline
$m_H$(GeV)& & & & &$\Delta\kappa$ & & & & &\\
 &-1.0&-0.6&-0.3&0.0&0.2&0.3&0.4&0.5&0.6&0.7~~~~~~\\
           & & & & & & & & & &~~~~~~\\[-2.5mm]
\hline
& & & & & & & & & &~~~~~~\\[-2.5mm]
 115 & 9 & 8 & 5 & 4 & 5 & 6 & 8 &$-$&$-$&$-$~~~~~~\\
 130 & 9 & 8 & 5 & 4 & 5 & 6 & 8 &$-$&$-$&$-$~~~~~~\\
 200 & 9 & 8 & 5 & 4 & 5 & 7 & 9 & 13 &$-$&$-$~~~~~~\\
 300 & 9 & 7 & 5 & 4 & 6 & 9 & 12 & 17 & 24 & 32~~~~~~\\[2mm]
\hline\hline
$pp\to ZZjj\to l^+l^-\nu\bar{\nu} jj$ & & & & & & & & &\\
\hline\hline
$m_H$(GeV)& & & & &$\Delta\kappa$ & & & & &\\
 &-1.0&-0.6&-0.3&0.0&0.2&0.3&0.4&0.5&0.6&0.7~~~~~~\\
           & & & & & & & & & &~~~~~~\\[-2.5mm]
\hline
& & & & & & & & & &~~~~~~\\[-2.5mm]
 115 & 5 & 4 & 2 & 0 & 1 & 3 & 5 &$-$&$-$&$-$~~~~~~\\
 130 & 5 & 4 & 2 & 0 & 1 & 3 & 5 &$-$&$-$&$-$~~~~~~\\
 200 & 5 & 4 & 2 & 0 & 1 & 3 & 5 & 8 &$-$&$-$~~~~~~\\
 300 & 5 & 4 & 2 & 0 & 1 & 3 & 6 & 9 & 14 & 20~~~~~~\\[2mm]
\end{tabular}
\end{table}

\begin{table}[h]
\null\noindent {\small Table VII. Decay branching ratio $B(H \to
\gamma \gamma)$, in the unit of $10^{-3}$, as a function of
$\Delta \kappa$ for various $m_H$.} \vspace*{4mm}
\begin{tabular}{cccccccccccc}
$m_H$(GeV)& & & & & &$\Delta \kappa$& & & & &\\
 &-0.4&-0.3&-0.2&-0.1&-0.05&0.0&0.05&0.1&0.2&0.3&0.4\\
\hline
 110&2.8 &2.4 &2.2 &2.0 &2.0 &1.9
   &1.9&1.8&1.8&1.7&1.7\\

 120&4.3 &3.2 &2.7 &2.4 &2.3  &2.2
   &2.1 &2.1 &1.9 &1.8 &1.7\\

 130&6.7  &4.4  &3.2  &2.6 &2.4  &2.3
   &2.1 &2.0 &1.8 &1.6 &1.5\\
\end{tabular}
\end{table}

\begin{table}[h]
\tabcolsep 2pt
\null\noindent
{\small Table VIII. Decay branching ratio 
$B(\kappa) \equiv B(H \to WW^*)$ as a function of
$|\kappa|$ for various $m_H$.}
\vspace*{4mm}
\begin{tabular}{cccccccccccc}
$m_H$(GeV)& & & & & &$|\kappa|$& & & & &\\
 &0.6&0.7&0.8&0.9&0.95&1.0&1.05&1.1&1.2&1.3&1.4\\
\hline
110&0.013 &0.018 &0.024 &0.03 &0.033 &0.037
   &0.040&0.044&0.052&0.061&0.069\\
120&0.048 &0.064 &0.082 &0.10 &0.11  &0.12
   &0.13 &0.14 &0.17 &0.19 &0.21 \\
130&0.12  &0.16  &0.20  &0.24 &0.26  &0.28
   &0.30 &0.32 &0.35 &0.39 &0.42 \\
150&0.48  &0.55  &0.60  &0.65 &0.67  &0.68
   &0.70 &0.71 &0.74 &0.76 &0.78\\
170&0.95  &0.95  &0.96  &0.96 &0.96  &0.97
   &0.97 &0.97 &0.97 &0.97 &0.97\\[2mm]
\end{tabular}
\end{table}

\tabcolsep 3pt

\begin{table}
\tabcolsep 3pt
\null\noindent {\small Table IX-A. Number of events at the LHC, with an
integrated luminosity of 300 ${\rm fb}^{-1}$, for 
$pp\to$ $W^\pm W^\pm$ $\to l^\pm \nu l^\pm \nu jj$
($l^{\pm}=e^{\pm}~{\rm or}~\mu^{\pm}$) in the linearly realized
effective Lagrangian with various values of $m_H$ and
$f_W/\Lambda^2$. The values of $N_S$/$\sqrt{N_S+N_B}$ are also shown
in the parentheses.}
\begin{tabular}{cccccccccccc}
$pp\to W^+W^+jj$&$\to l^+\nu l^+\nu jj$ & & & & & & & & & &  \\
\hline\hline
$m_H$(GeV)& & & & & &$f_W/\Lambda^2$ & (TeV$^{-2})$& & & &  \\
 &-4.0&-3.0&-2.0&-1.4&-1.0
 &0.0
 &0.85&1.2&2.0&3.0&4.0\\
\hline
 115&117(9.4)&72(6.7)&38(3.7)&26(2.2)&20(1.1)
 &15
 &20(1.1)&25(2.0)
 &42(4.2)&78(7.1) &129(10)\\
 130&118(9.5)&72(6.7)&38(3.7)&26(2.2)&20(1.1)
 &15
 &20(1.1)&25(2.0)
 &42(4.2)&78(7.1) &130(10)\\
 200&119(9.5)&73(6.8)&38(3.7)&26(2.2)&20(1.1)
 &15
 &20(1.1)&25(2.0)
 &42(4.2)&79(7.2) &132(10)\\
 300&121(9.6)&75(6.9)&39(3.8)&27(2.3)&21(1.3)
 &16
 &21(1.3)&26(2.2)
 &43(4.3)&80(7.3) &134(10)\\
\hline\hline
$pp\to W^-W^-jj$&$\to l^-\bar{\nu}l^-\bar{\nu}jj$ & & & & & & & & \\
\hline\hline
$m_H$(GeV)& & & & & &$f_W/\Lambda^2$ & (TeV$^{-2})$& & & &  \\
 &-4.0&-3.0&-2.0&-1.4&-1.0
 &0.0
 &0.85&1.2&2.0&3.0&4.0\\
\hline
 115&23&14&7&5&4
 &3
 &4&5
 &8&15 &25\\
 130&23&14&7&5&4
 &3
 &4&5
 &8&15 &25\\
 200&23&14&7&5&4
 &3
 &4&5
 &8&15 &26\\
 300&24&15&8&5&4
 &4
 &4&5
 &8&16 &26
\end{tabular}
\end{table}
\null\vspace{0.2cm}

\begin{table}
\tabcolsep 3pt
\null\noindent {\small Table IX-B. Number of events at the LHC, with an
integrated luminosity 300 ${\rm fb}^{-1}$, for 
$pp\to$ $W^\pm W^\pm$ $\to l^\pm \nu l^\pm \nu jj$
($l^{\pm}=e^{\pm}~{\rm or}~\mu^{\pm}$) in the linearly realized
effective Lagrangian with various values of $m_H$ and
$f_{WW}/\Lambda^2$. The values of $N_S$/$\sqrt{N_S+N_B}$ are
also shown in the parentheses.}
\begin{tabular}{cccccccccccc}
$pp\to W^+W^+jj$&$\to l^+\nu l^+\nu jj$ & & &  & & & & &  \\
\hline\hline
$m_H$(GeV)& & & & &$f_{WW}/\Lambda^2$  (TeV$^{-2})$& & & &  \\
 &-4.0&-3.0&-2.2&-1.6
 &0.0
 &1.6&2.2&3.0&4.0\\
\hline
 115&47(4.7)&33(3.1)&25(2.0)&19(0.9)
 &15
 &20(1.1)
 &26(2.2)&33(3.1) &48(4.8)\\
 130&48(4.8)&33(3.1)&25(2.0)&19(0.9)
 &15
 &20(1.1)
 &26(2.2)&34(3.1) &49(4.9)\\
 200&49(4.9)&34(3.3)&25(2.0)&19(0.9)
 &15
 &20(1.1)
 &26(2.2)&35(3.4) &50(4.9)\\
 300&51(5.0)&35(3.4)&26(2.2)&20(1.1)
 &16
 &21(1.3)
 &27(2.3)&36(3.5) &52(5.1)\\
\hline\hline
$pp\to W^-W^-jj$&$\to l^-\bar{\nu}l^-\bar{\nu}jj$ & & & & & & & & \\
\hline\hline
$m_H$(GeV)& & & & &$f_{WW}/\Lambda^2$  (TeV$^{-2})$& & & &  \\
 &-4.0&-3.0&-2.2&-1.6
 &0.0
 &1.6&2.2&3.0&4.0\\
\hline
 115&9&6&5&4
 &3
 &4
 &5&6 &9\\
 130&9&6&5&4
 &3
 &4
 &5&7 &10\\
 200&10&7&5&4
 &3
 &4
 &5&7 &10\\
 300&10&7&5&4
 &4
 &4
 &5&7 &10

\end{tabular}
\end{table}
\null\vspace{0.2cm}

\begin{table}
\null\noindent {\small Table X. Number of events at the LHC, with an
integrated luminosity of 300 fb$^{-1}$, for 
$pp\to  W^+W^-jj$ $\to l^+\nu l^-\nu jj$ ($l^\pm=e^\pm$ or
$\mu^\pm$) in the linearly realized effective Lagrangian with
various values of $m_H$ and $f_{W}/\Lambda^2$.}
\begin{tabular}{ccccccccccc}
$m_H$(GeV) & & & & & $f_W/\Lambda^2$ (TeV$^{-2}$)& & & &  \\
&-4.0&-3.0&-2.0&-1.0
&0.0
&1.0&2.0&3.0&4.0\\
\hline 115&35&21&10&5
 &4
 &6&12&23&38\\
130&35&21&10&5
&4
&6&12&23&38\\
200&36&21&10&5
&4
&6&12&23&39\\
300&37&22&11&5
&4
&6&13&24&40
\end{tabular}
\end{table}

\begin{table}
\null\noindent {\small Table XI. Number of events at the LHC, with an
integrated luminosity 300 fb$^{-1}$, for
$pp\to$ $ ZZjj\to l^+l^- l^+l^- (l^+l^- \nu\bar\nu )jj$
($l^\pm=e^\pm$ or $\mu^\pm$) in the linearly realized effective
Lagrangian with various values of $m_H$ and $f_{W}/\Lambda^2$.}
\begin{tabular}{ccccccccccc}
$pp\to ZZjj$$\to l^+l^-\nu\bar{\nu}jj$& & & & & & & & & & \\
\hline\hline
$m_H$(GeV)& & & & &$f_W/\Lambda^2$ (TeV$^{-2}$)& & & & \\
 &-4.0&-3.0&-2.0&-1.0
 &0.0
 &1.0&2.0&3.0&4.0\\
\hline 115&10&6&2&1
 &0
 &1&3&6&11\\
130&10&6&2&1
 &0
 &1&3&6&11\\
200&10&6&2&1
 &0
 &1&3&6&11\\
300&10&6&2&1
&0
&1&3&6&12\\
\hline\hline
$pp\to ZZjj$$\to l^+l^-l^+l^-jj$& & & & & & \\
\hline\hline
$m_H$(GeV)& & & & &$f_W/\Lambda^2$ (TeV$^{-2}$)& & & & \\
 &-4.0&-3.0&-2.0&-1.0
 &0.0
 &1.0&2.0&3.0&4.0\\
\hline 115&9&7&5&4
 &4
 &5&6&8&10\\
130&9&7&5&4
 &4
 &5&6&8&10\\
200&9&7&5&4
 &4
 &5&6&8&10\\
300&9&7&5&4
 &4
 &5&6&8&11

\end{tabular}
\end{table}
\null\vspace{0.2cm}

\begin{table}
\null\noindent {\small Table XII. Number of events at the LHC, with an
integrated luminosity 300 fb$^{-1}$, for 
$pp\to  ZW^\pm jj$ $\to  l^\pm \nu l^+l^- jj$ ($l^\pm=e^\pm$ or
$\mu^\pm$) in the linearly realized effective Lagrangian with
various values of $m_H$ and  $f_{W}/\Lambda^2$.}
\begin{tabular}{ccccccccccc}
$pp\to ZW^+ jj\to l^+l^-l^+\nu jj$ & & & & & & & & & &   \\
\hline\hline
$m_H$(GeV) & & & & &$f_W/\Lambda^2$ (TeV$^{-2})$& & & & \\
&-4.0&-3.0&-2.0&-1.0
&0.0
&1.0&2.0&3.0&4.0\\
\hline 115&10&7&4&3
&2
&3&5&8&12\\
130&10&7&4&3
&2
&3&5&8&12\\
200&10&7&4&3
&2
&3&5&8&12\\
300&10&7&4&3
&2
&3&5&8&12\\
\hline\hline
$pp\to ZW^- jj\to l^+l^-l^-\bar{\nu}jj$ & & & & & & & & & &   \\
\hline\hline
$m_H$(GeV) & & & & &$f_W/\Lambda^2$ (TeV$^{-2})$ & & & & \\
 &-4.0&-3.0&-2.0&-1.0
 &0.0
 &1.0&2.0&3.0&4.0\\
\hline 115&4&2&1&1
&1
&1&1&2&4\\
130&4&2&1&1
&1
&1&1&2&4\\
200&4&2&1&1
&1
&1&1&2&4\\
300&4&2&1&1
&1
&1&1&2&4

\end{tabular}
\end{table}

\begin{table}
\null\noindent {\small Table\,XIII. Summary of the
$2\sigma$ constraints on the anomalous $HVV$ coupling
$\kappa$ ($\equiv 1+ \Delta\kappa$) from the dimension-3 operator
in the nonlinear Higgs sector studied in Sec.\,III.}
\begin{tabular}{ccc}
Types of constraints&results& Places in the text
\\[1.5mm]
\hline Precision EW data&regions shown in Figs.\,2 and 3, Table\,I
&Eq.\,(7)
\\[1.5mm]
Unitarity (at $\sqrt{s}$=2 TeV)&$0.5<|\kappa|<1.3$& Eq.\,(11)
\\[1.5mm]
$W^+W^+$ scattering&$-0.3<\Delta\kappa<0.2$& Eq.\,(13)
\\[1.5mm]
$HV$ production ($m_H=120$ GeV)& $0\leq |\kappa|\leq 1.6$&
Eq.\,(14)
\\[1.5mm]
Higgs width ($m_H=200\-- 300$ GeV)& $0.8\leq|\kappa|\leq
1.2$& Eq.\,(15)
\\[1.5mm]
B($H \to \gamma \gamma$)& Table\,VII & Sec.\,VC
\\[1.5mm]
Higgs resonance ($m_H=120$ GeV)& $0.88\leq R\leq 1.12~~[R\equiv
\kappa^2B(\kappa)/B(\kappa=1)]$, Table\,VIII & Eq.\,(16)
\\[1.5mm]
LC (500\,GeV, $1 {\rm ab}^{-1}$) & $|\Delta \kappa| \leq 0.3\%$ 
& Sec.\,VE
\\[1.5mm]
\end{tabular}
\end{table}

\begin{table}
\null\noindent {\small Table XIV. Summary of the $2\sigma$
constraints on the anomalous couplings
$f_n/\Lambda^2$ (in units of TeV$^{-2}$) associated
with the dimension-6 operators [cf. Eqs.\,(17)--(19)]
and the related form factors $g^{(1)}_{HVV}$ and
$g^{(2)}_{HVV}$ (in units of TeV$^{-1}$) [cf. Eq.\,(22)] in the linearly
realized Higgs sector studied in Sec.\,IV.}
\vspace*{2mm}
\renewcommand{\baselinestretch}{1.8}
\begin{tabular}{ccc}
types of constraints&results&places in the text\\
\hline
 precision EW data
 &two-parameter fit at tree level: regions shown in Fig. 4, with 
  ~~~~~~&Fig.
 4\\
 & $-0.05 < \frac{f_{\Phi,1}}{\Lambda^2}< 0.02$,~~~~~~~~~~
   $-0.10 < \frac{f_{BW}}{\Lambda^2}< 0.05$. & 
\\
 &one-parameter fit at one-loop level:~~~~~~~~~~~~~~~~~~~~~~~~~~~~~~~~~~~~~~~~& \\
 &$-6\leq \frac{f_{WWW}}{\Lambda^2}\leq 3$,~~~~$-6\leq
\frac{f_W}{\Lambda^2}\leq 5$,~~~~
$-4.2\leq\frac{f_B}{\Lambda^2}\leq
2.0$,& \\
 &$-5.0\leq\frac{f_{WW}}{\Lambda^2}\leq 5.6$,~~~~
$-17\leq\frac{f_{BB}}{\Lambda^2}\leq 20. $&Eq. (23)\\
triple gauge coupling data& $-31\leq\frac{(f_W+f_B)}{\Lambda^2}\leq 68$ 
(for $f_{WWW}=0$),& \\
 &$-41\leq\frac{f_{WWW}}{\Lambda^2}\leq 26$ (for
$f_W+f_b=0$). &Eq. (24)\\
LEP2 Higgs searches& $-7.5\leq \frac{f_{WW(BB)}}{\Lambda^2}\leq
18. $&Eq. (25)\\
unitarity requirement &$|\frac{f_b}{\Lambda^2}|\leq
24.5$,~~~~ $|\frac{f_W}{\Lambda^2}|\leq 7.8$,~~~~
$|\frac{f_{WWW}}{\Lambda^2}|\leq 7.5$,& \\
  (at $\sqrt{s}$=2 TeV)
  &$-160\leq |\frac{f_{BB}}{\Lambda^2}|\leq 197$,~~~~
$|\frac{f_{WW}}{\Lambda^2}|\leq 39.2. $&Eq. (27)\\
$W^+W^+$ scattering&$1\sigma$~~~~~~~~~~~~~~~~~~~~~~~~~~~~~~~~~~~~~~~$2\sigma$~~~~ & \\
 &$-1.0<~\frac{f_W}{\Lambda^2}< 0.85$,~~~~~~~~~~~~~~~~~~
 $-1.4<\frac{f_W}{\Lambda^2}\leq 1.2. $
 ~~~~~~&Eq. (29)\\
 &$-1.6<\frac{f_{WW}}{\Lambda^2}<1.6$,~~~~~~~~~~~~~~~~~~
 $-2.2\leq\frac{f_{WW}}{\Lambda^2}<2.2$.~~~~&
 Eq. (30) \\
 &or~~~~~~~~~~~~~~~~~~~~~~~~~~~~~~~~~~~~~~~~~~~~~~~~~~~~~~~~~~~~~~~~~~~~~~~~~~~~~~~~~~~~~~~~~& \\
 &$-0.026< g^{(1)}_{HWW}< 0.022$,
 ~~~~~~~~~~$-0.036< g^{(1)}_{HWW}\leq 0.031$. & \\
 &$-0.026< g^{(1)}_{HZZ}< 0.022$,
 ~~~~~~~~~~$-0.036< g^{(1)}_{HZZ}\leq 0.031$. & \\
 &$-0.014< g^{(1)}_{HZ\gamma}< 0.012$,
 ~~~~~~~~~~$-0.020< g^{(1)}_{HZ\gamma}\leq 0.017$. & \\
 &$-0.083< g^{(2)}_{HWW}< 0.083$,
 ~~~~~~~~~~$-0.11~\leq g^{(2)}_{HWW}< ~0.11$. & \\
 &$-0.032< g^{(2)}_{HZZ}< 0.032$,
 ~~~~~~~~~~$-0.044\leq g^{(2)}_{HZZ}< 0.044$. & \\
 &$-0.018< g^{(2)}_{HZ\gamma}< 0.018$,
 ~~~~~~~~~~$-0.024\leq g^{(2)}_{HZ\gamma}< 0.024$. &
 Eq. (31)\\
 $HVV$ coupling at & & 
  \\
LC (500\,GeV, $1~ {\rm ab}^{-1}$) & 
$|\frac{f_{\Phi,2}}{\Lambda^2}| < 0.2 $ & Sec. VIA
 \\
\end{tabular}
\end{table}
\newpage

\newpage
\begin{figure}[ht]
\centerline{\epsfig{figure=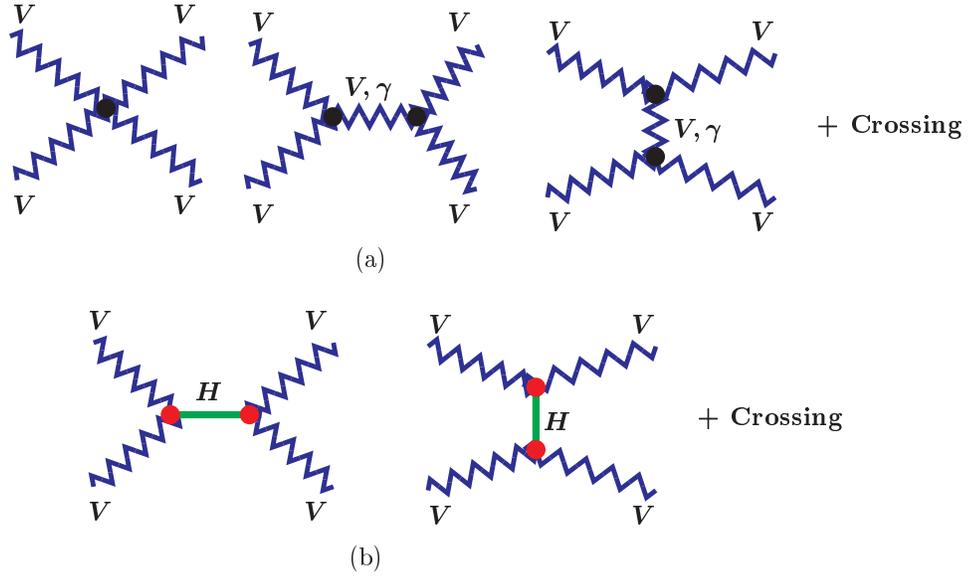,width=19cm,height=9cm}}
\null\noindent \caption{Illustration of Feynman diagrams for $VV$
scatterings in the SM: (a) diagrams contributing to $T(V,\gamma)$,
(b) diagrams contributing to $T(H)$.} \label{Fig.1}
\end{figure}

\null\vspace{0.4cm}
\begin{figure}[t]
\vspace*{-16mm}
\centerline{\epsfig{figure=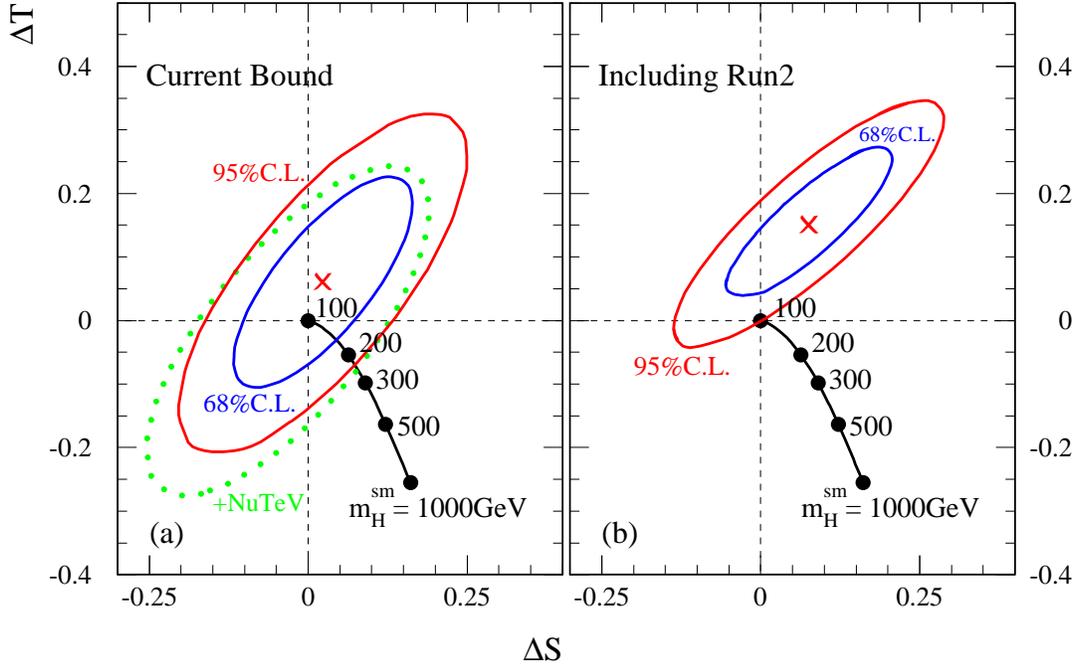,height=11cm}}
\vspace{-0.2cm} \caption{$\Delta S -\Delta T$ contours (a) from
the current precision electroweak data, and (b) from including the
expected Tevatron Run-2 measurement of $m_W$ and $m_t$ (assuming
the current central values of $m_W$ and $m_t$ with an error of
$20$\,MeV and $2$\,GeV, respectively). Here, we have set $m_H^{\rm
ref}=100$\,GeV and $\Delta U=0$ in the fit. } \label{Fig:Fig1}
\end{figure}

\vspace{0.4cm}
\begin{figure}[t]
\vspace*{-13mm}
\hspace*{1.2cm}
\centerline{\epsfig{figure=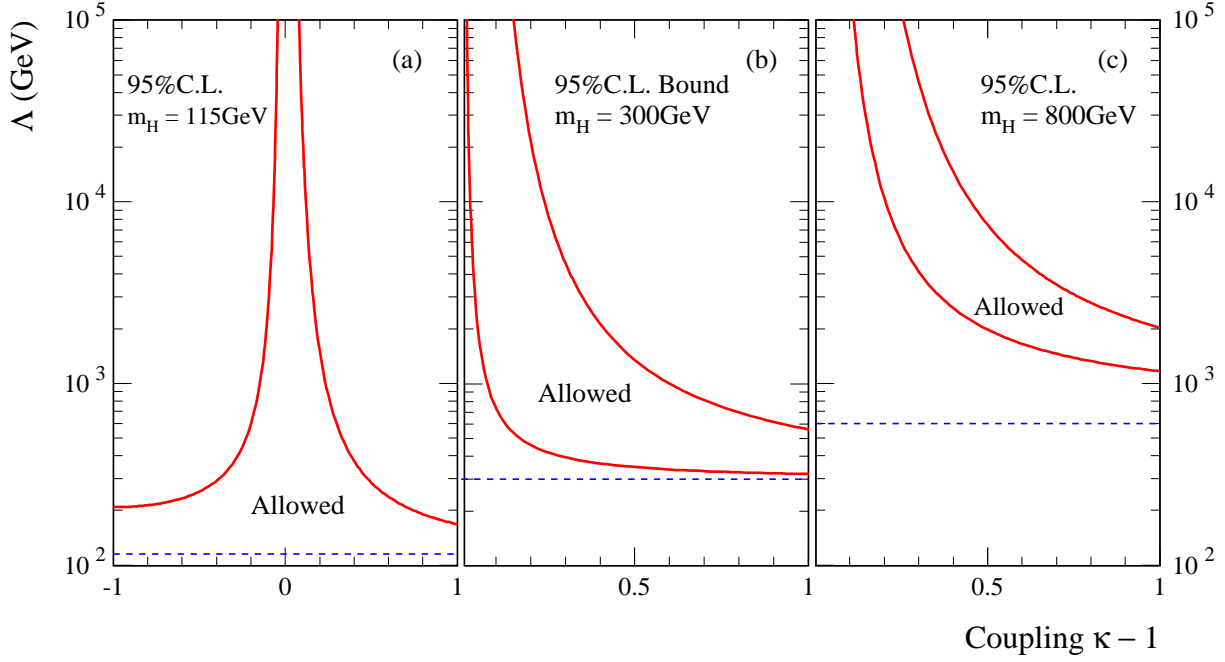,height=10.5cm}}
\vspace*{-10mm} \vspace*{1cm} \caption{Constraints on the new
physics scale $\Lambda$ as a function of the anomalous coupling
$\Delta \kappa \equiv \kappa -1$. The regions below the solid
curves and above the dashed lines [(a)] or between the two
solid curves [(b)-(c)] are allowed at the $95\%$ C.L. The dashed
lines indicate the value of $m_H^{\rm ref}=m_H$.} \label{Fig:Fig2}
\end{figure}

\begin{figure}[h]
\epsfxsize=11.3cm \centerline{\epsffile{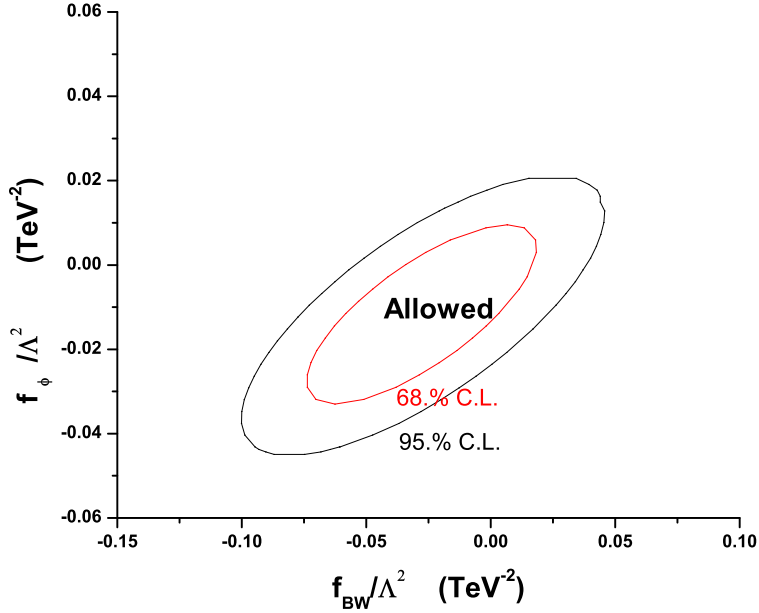}}
\caption{The $68\%$ and $95\%$ C.L. bounds on $f_{BW}/\Lambda^2$ and
$f_{\Phi,1}/\Lambda^2$ (in units of TeV$^{-2}$) from the tree
level formulas of $\Delta S$ and $\Delta T$ [40]  
and the
$\Delta S$-$\Delta T$ bounds given in Fig. 2(a).}
\label{Fig. 4}
\end{figure}

\begin{figure}[h]
\epsfxsize=10cm \centerline{\epsffile{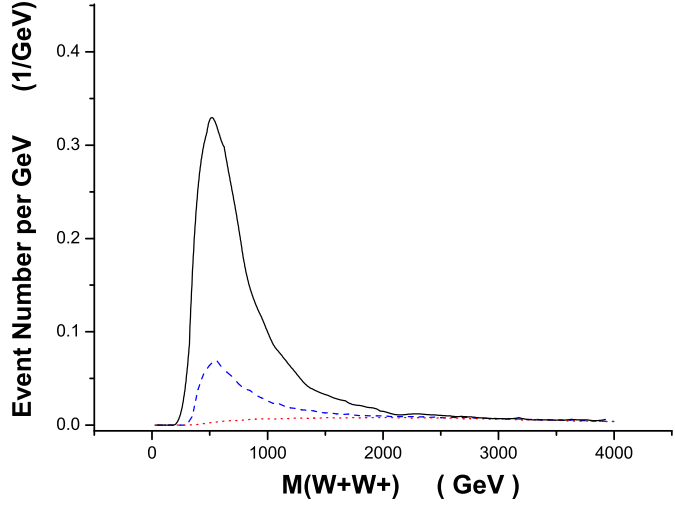}} \null\noindent
\vspace{-0.8cm} \caption{Invariant mass distributions of the
$W^+W^+$ pairs produced at the LHC for $m_H=115$ GeV with $f_W/\Lambda^2=$
$f_B/\Lambda^2$=5 TeV$^{-2}$. The solid curve is the result from
the complete tree level calculation; the dashed curve is the result
from the EWA calculation with the exact $W^+_LW^+_L \to W^+_LW^+_L$
amplitude; the dotted curve is the result from the EWA
calculation with the asymptotic formula [49] for
the $W^+_LW^+_L \to W^+_LW^+_L$ scattering amplitude.}
 \label{MWW}
\end{figure}


\begin{figure}[ht]
\centerline{\epsfig{figure=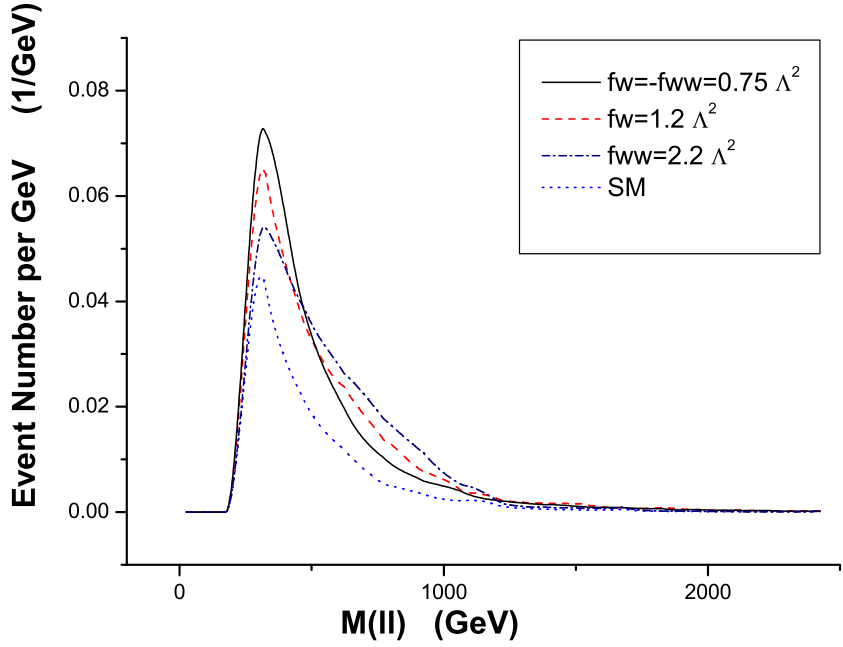,height=10cm}}
\null\vspace{-0.8cm} \null\noindent \caption{Invariant mass
distribution of the di-leptons from the decay of $W^+$ bosons
produced at the LHC via $pp \to W^+W^+jj$ for $m_H=115$ GeV with
$f_W/\Lambda^2=$
$-f_{WW}/\Lambda^2=0.75$~TeV$^{-2}$ [Eq. (\ref{f})] (solid line),
$f_W/\Lambda^2=1.2$~TeV$^{-2}$, $~f_{WW}/\Lambda^2=0$ [Eq.
(\ref{fW})] (dashed line), $f_{WW}/\Lambda^2=2.2$~TeV$^{-2}$,
$~f_W/\Lambda^2=0$ [Eq. (\ref{fWW})] (dashed-dotted line), and the
SM (dotted line). } \label{Mll}
\end{figure}
\end{document}